\documentclass[twocolumn]{article}
\usepackage[paper=a4paper,margin=2cm,includehead=false]{geometry}
\usepackage{fullpage}
\usepackage{amsmath}
\usepackage{amssymb}
\usepackage{blindtext}
\usepackage{ifthen}

\usepackage{pgfplots}
\usepgfplotslibrary{statistics}
\pgfplotsset{compat=newest}
\pgfplotsset{minor grid style={dashed,very thin, color=blue!15}}
\pgfplotsset{major grid style={very thin, color=black!30}}

\pgfkeys{/pgf/number format/.cd,
fixed,
fixed zerofill=false,
set thousands separator={\,},
1000 sep in fractionals,
sci E,
}

\newenvironment{experimentfiguressmall}{\begin{figure}[ptb]}{\end{figure}}
\newenvironment{experimentfigure}{\begin{figure*}}{\end{figure*}}

\makeatletter
\tikzset{nomorepostaction/.code=\let\tikz@postactions\pgfutil@empty}
\makeatother

\pgfplotsset{
	automatically generated axis/.style={
		height=85pt,
		width=135pt,
		scaled ticks=false,
		xticklabel style={font=\tiny,/pgf/number format/.cd, fixed,/tikz/.cd},
		yticklabel style={font=\tiny,/pgf/number format/.cd, fixed,/tikz/.cd},
		x label style={at={(ticklabel cs:0.5, -5pt)},name={x label},anchor=north,font=\scriptsize},
		y label style={at={(ticklabel cs:0.5, -5pt)},name={y label},anchor=south,font=\scriptsize},
	},
	automatically generated symbolic/.style={
		height=105pt,
		width=500pt,
		xticklabel style={font=\tiny,rotate=90},
		yticklabel style={font=\tiny,/pgf/number format/.cd, fixed,/tikz/.cd},
		x label style={at={(ticklabel cs:0.5, -5pt)},name={x label},anchor=north,font=\scriptsize},
		y label style={at={(ticklabel cs:0.5, -5pt)},name={y label},anchor=south,font=\scriptsize},
	},
	first kind/.style={
		legend style={font=\scriptsize,fill=none},
		legend columns=6,legend cell align=left,
	},
	posterior kind/.style={
		legend style={draw=none},
	},
	bar legend/.style={
       legend style={font=\scriptsize,fill=none},
       legend columns=4,legend cell align=left,
	   area legend,
	}
}
\tikzset{
	automatically generated plot/.style={
		/pgfplots/error bars/error bar style={ultra thin,solid,opacity=0.25},
		/tikz/mark options={solid},
	},
	automatically generated bar plot/.style={
	},
	automatically generated boxplot/.style={
	},
	temporal plot/.style={
		mark repeat=100,
		mark phase=5,
		/pgfplots/error bars/x dir=none,
		/pgfplots/error bars/y dir=none,
	},
}

\tikzset{
	error pattern/.style 2 args={
		/pgfplots/error bars/draw error bar/.code 2 args={%
			\pgfmathsetlengthmacro{\halfwidth}{.5*\pgfkeysvalueof{/pgf/bar width}}
			\path[draw=none,line width=0pt,xshift={\pgfplotbarshift},fill=#1,postaction={pattern=#2,pattern color=white}] ##1 -- ++(\halfwidth,0) |- ##2 -- ++(-\halfwidth,0) |-cycle;
		},%
	},
	swap bar/.style 2 args={
		automatically generated bar plot,
		fill=#1!20,
		postaction={pattern=#2},
		error pattern={#1!50!black}{#2},
	},
}
\tikzset{xxxxRandomxxxxxbar/.style={swap bar={red}{horizontal lines}}}
\tikzset{xxxxRowsxxxxxbar/.style={swap bar={green}{grid}}}
\tikzset{xxxxRRxswitchesxxxxxbar/.style={swap bar={blue!90!white}{crosshatch}}}
\tikzset{xxxxRVIIIxRIVxRIIxxxxxbar/.style={swap bar={black}{dots}}}
\tikzset{xxxxLxtilesxxxxxbar/.style={swap bar={violet}{north east lines}}}
\tikzset{xxxxDiagonalsxxxxxbar/.style={swap bar={orange}{vertical lines}}}
\tikzset{xxxxRandomxSwitchxSelectionxxxxxbar/.style={swap bar={lightgray}{horizontal lines}}}

\def\xxxxRandomxxxxxtext{Random Endpoint}
\def\xxxxDiagonalsxxxxxtext{Diagonal Selection}
\def\xxxxRVIIIxRIVxRIIxxxxxtext{Rectangular Tessellation}
\def\xxxxLxtilesxxxxxtext{L-shape Tessellation}
\def\xxxxRandomxSwitchxSelectionxxxxxtext{Random Switch}
\def\xxxxRowsxxxxxtext{Row Selection}
\def\xxxxRRxswitchesxxxxxtext{Full-Spread}

\usepackage[utf8]{inputenc}%
\usepackage{algorithm}
\usepackage{algorithmic}
\usepackage{subcaption}
\usetikzlibrary{positioning}%
\usetikzlibrary{decorations.text}%
\usetikzlibrary{calc}%
\usetikzlibrary{arrows, decorations.markings}%
\usetikzlibrary{matrix}%
\usetikzlibrary{patterns}%
\usetikzlibrary{external}%
\usetikzlibrary{arrows.meta}%
\usetikzlibrary{intersections}%
\usepackage{url}
\usepackage{pgfplotstable}
\newlength{\heatmapcellsize}%
\pgfplotstableset{
	heatmap/.style={
		every head row/.style={output empty row},
		font=\scriptsize,
		postproc cell content/.code={%
			\pgfkeysalso{@cell content=%
				\rule[-.2\heatmapcellsize]{0pt}{\heatmapcellsize}%
				\cellcolor{black!##1}%
				\pgfmathtruncatemacro\number{##1}\ifnum\number>50\color{white}\fi%
				\makebox[\heatmapcellsize]{\number}}%
		},
	}
}

\newtheorem{definition}{Definition}
\usepackage{colortbl}%
\DeclareMathOperator{\rem}{rem}
\DeclareMathOperator{\quo}{quo}
\DeclareMathOperator{\PB}{\textit{PB}}
\usepackage{environ}
\usepackage{booktabs}
\NewEnviron{lesson}{\bgroup\vspace{1em plus 1em minus .5em}\noindent\begin{tikzpicture}
\node[text width=.95\columnwidth,draw,ultra thick,fill=blue!10,rounded corners=2pt] {\BODY};
\end{tikzpicture}\vspace{.5em plus .5em minus .4em}\egroup}
\newcommand{\prettynumber}[1]{\pgfmathprintnumber[
	fixed,
	fixed zerofill=false,%
	set thousands separator={\hspace{1pt plus 1pt minus .5pt}},
	1000 sep in fractionals,
	sci E,%
]{#1}}

\title{Resource Allocation in HyperX Networks}
\author{Alejandro Cano$^\dagger$, Cristóbal Camarero$^\dagger$, Carmen Martínez$^\dagger$, Ramón Beivide$^{\dagger,\ast}$\\
	$^\dagger$\textit{Universidad de Cantabria}, SPAIN\\
	$^\ast$\textit{Barcelona Supercomputing Center}, SPAIN\\
	\small\{alejandro.cano, cristobal.camarero, carmen.martinez,ramon.beivide\} @unican.es
}
\begin{document}
	\maketitle
	\thispagestyle{plain}
	\pagestyle{plain}

	\begin{abstract}
		As high-performance computing systems scale in size and complexity, efficient resource management is essential to minimize communication overhead.
		The HyperX is a richly connected, low-diameter network that offers a scalable and cost-effective alternative to traditional topologies.
		However, resource allocation in HyperX remains underexplored, and strategies designed for networks like Torus, Fat-tree, or Dragonfly do not directly transfer.
		In this work, we propose and formalize several resource allocation strategies for HyperX networks, categorized into linear, geometric, and stochastic functions.
		We characterize these strategies theoretically by analyzing their topological properties, including dilation, convexity, and partition bandwidth.
		Furthermore, we conduct an exhaustive experimental evaluation using synthetic traffic and application communication kernels to assess the impact of these strategies on performance under different routing algorithms.
		Our results indicate that partition bandwidth and switch locality are decisive factors in mitigating interferences.
		Notably, the Diagonal allocation strategy, which is not convex, consistently outperforms traditional approaches in most scenarios.
		Finally, we provide a set of lessons learned to guide the implementation of resource allocation policies in HPC systems based on HyperX networks.

	\end{abstract}

	\section{Introduction}

Efficient resource management has become a critical concern for HPC system architects, administrators, and users.
Current parallel applications often require thousands of processing cores, making the communication overhead between compute nodes, or network endpoints, a key performance bottleneck.
	To fully exploit the computational power of these systems, minimizing communication costs is essential.

	In practice, HPC systems receive a diverse set of jobs, each representing an application composed of multiple processes or tasks that must be executed across the system’s available endpoint computers.
	The process of assigning computing resources to applications is known as job allocation or \textit{resource allocation}.
	Subsequently, the task mapping assigns each task of the application to a specific compute node within the selected set.
	Both steps are critical to minimizing communication overhead and improving the overall performance of the system.

	In most HPC systems, job allocation is handled by workload managers such as SLURM, which queue and assign jobs to available resources based on system policies.
	While such tools provide the framework for managing system workload, the underlying allocation strategies must account for the characteristics of the interconnection network, as they indirectly determine how network resources are utilized.
	For example, if all the endpoints allocated to an application are connected to the same switch, the application’s communication will remain local and avoid traversing any network link.
	However, if the application requires more endpoints than those available under a single switch, the job allocator must select a set of endpoints from different switches to host the application.
	Depending on the physical location of these switches, the application's traffic may traverse different network paths, leading to variations in latency and available bandwidth.
	For this reason, resource allocation has long been an active area of research across a wide range of networks.

\input{figures/topologias}

	Current HPC systems are increasingly adopting modern low-diameter topologies such as Dragonfly~\cite{Kim_dgfly_ISCA}, Dragonfly+~\cite{dragonfly_plus}, and HyperX~\cite{HyperX}, which offer greater scalability and lower cost than Fat-trees.
	Representations of some of these topologies can be seen in Figure~\ref{fig:topologias}.
	In them, circles represent switches and lines that join them are physical cables.
Logically, these switches and cables form a graph.
The rounded rectangles are the endpoints, each one connected to a single switch.
The \textsl{diameter} of a network is the number of switch-to-switch links that are necessary to traverse to connect any source-destination pair, using minimal paths.

	Nowadays, 8 out of the 10 most powerful supercomputers in the TOP500 list employ one of these low diameter networks~\cite{TOP500}. Resource allocation in such networks presents challenges distinct from those addressed by traditional approaches for low-radix topologies, such as tori, or full-bisection bandwidth fabrics, like Fat-trees.
	While modern topologies exhibit low distances between endpoints, they are often constrained by limited bisection bandwidth.

	In this work, we focus on the \textit{HyperX} topology—a richly connected, low-diameter network that has gained traction in HPC system design but remains underexplored in the context of resource allocation. Our primary objective is to analyze how different resource allocation strategies affect application performance when executed on systems based on a HyperX network.
	The main contributions of the paper are:

	\begin{itemize}

		\item A theoretical model for the different possibilities of resource allocation in HyperX networks based on mathematical functions with different foundations: geometric, algebraic, and random.

		\item A complete characterization of the most critical properties of the applications derived from the aforementioned theoretical modeling, namely: dilation, convexity and bandwidth.

		\item An exhaustive evaluation through simulation of the proposed techniques.

		\item A comprehensive list of outcomes from which architects, administrators, and programmers of HPC systems based on HyperX topologies could benefit.

	\end{itemize}

The remainder of this paper is organized as follows. Section~\ref{sec:HyperX} introduces the fundamentals of HyperX networks, including topology, routing, and scalability. Section~\ref{sec:background} presents the motivation of this work and reviews resource allocation strategies in well-known topologies such as Fat-tree, Torus, and Dragonfly. Section~\ref{sec:resource_selection} formulates the proposed resource allocation management using mathematical functions.
Their fundamental properties are analyzed in Section~\ref{sec:analysis}. The proposed approaches are then experimentally validated. Section~\ref{sec:application_analysis} describes the experimental methodology, while Section~\ref{sec:experiments} presents the obtained results. Finally, Section~\ref{sec:outcomes} summarizes the main lessons learned of the complete study. Section~\ref{sec:discussion} concludes the paper and briefly discusses the extension of the techniques presented to higher-dimensional HyperX networks.

	\section{The HyperX Network} \label{sec:HyperX}

This section introduces HyperX networks by defining their topology and the routing mechanism employed, and compares their scalability to other high-degree low-diameter networks.

	\subsection{Topology}

	The \textit{HyperX} topology is a low-diameter, high-radix interconnection network proposed as a scalable alternative to traditional networks such as Fat-trees. It has been adopted in systems like Intel's PIUMA~\cite{PIUMA}, Cray's Cascade~\cite{Cascade} and in \cite{FT-hyperx}, underscoring its relevance in current and forthcoming HPC architectures.

The \textsl{HyperX networks} are those networks having a Hamming graph as the topology of their switches \cite{HyperX}.
	The \textsl{Hamming graph} is the Cartesian graph product of complete graphs.
	A complete graph has a link between every pair of vertices.

	The name of Hamming graph refers to the fact that the graph distance, that is the count of hops in a minimal path, coincides with the Hamming distance, used in Coding theory.
	This name has appeared in Graph Theory from at least 1974 \cite{Biggs,Mulder}.
	As a computer network, it was considered in \cite{Bhuyan} under the name of Generalized Hypercube.
	In that paper the authors noted that the Butterfly network could be understood as an unflattening of the Hamming graph.
	Later, in \cite{LaForge} the same graphs were considered as computer networks, under the pretense of enabling interstellar travel.
	In \cite{Kim_flat_ISCA} these networks received the name of Flattened Butterflies.
	Then, in \cite{HyperX} it received a flashier name: HyperX, which is the one employed in Intel's PIUMA \cite{PIUMA}.
	It is to be noticed that previous definitions differ in aspects such as enforcing all sides to be equal, the number of endpoints per switch, whether to use parallel links for some edges, or if links are electrical or optical.

	Regardless of these details, a symmetric $q$D HyperX organizes its switches in a $q$-dimensional logical grid with each side having the same size, $n$.
	Each switch $S$ has an address of $q$ coordinates $(s_{q}, s_{q-1}, \dotsc, s_{2}, s_{1})$, with $s_{i}\in\{0, 1, \dotsc, n-1\}$.
	Two switches are connected by an edge if their Hamming distance is 1, meaning that their addresses differ only in a single coordinate.

	Endpoint computers are connected to every switch.
	A direct network, as HyperX, is said to be well-balanced when, under uniform random traffic, every endpoint can inject data at full rate.
	A well-balanced $q$D HyperX of side $n$, requires $n$ local ports on each of its $n^{q}$ switches to connect endpoint computers, leading to a total of $n^{q+1}$ endpoints in the system.
	In addition, the switches require $q(n-1)$ network ports, $(n-1)$ of them per each of its $q$ dimensions, to implement the complete graphs that compose each dimension.
	The diameter in a $q$D HyperX is $q$.
	This just considers traversed network links, disregarding injection and ejection local connections.
	In the same way, the average distance between pairs of switches is $q - q/n$.
Note that this calculation, as well as all subsequent in the paper, is obtained assuming that every source may have itself as destination,
which simplify its mathematical expression and further manipulations.

	The total number of network wires is $q(n-1)n^{q}/2$.
	In terms of raw cost, each endpoint computer requires $1/n$ switches and $q(n-1)/2n$ wires.
	This is, the cost in wires per endpoint computer approaches $q/2$ from below.

\begin{figure}
    \centering
    \usetikzlibrary{backgrounds}
    \pgfdeclarelayer{background}
    \pgfsetlayers{background,main}
    \begin{tikzpicture}[
        scale=0.5,
        x=2.0cm, y=2.0cm,
        switch/.style={
            draw=black!80, circle, thick, fill=white,
            inner sep=0pt, minimum width=15pt,
            font=\scriptsize\sffamily
        },
        server/.style={
            draw=gray!60, rectangle, rounded corners=1pt, fill=gray!20,
            inner sep=0pt, minimum width=4pt, minimum height=2.5pt
        },
        cable/.style={draw=black, line width=0.5pt},
        serverlink/.style={draw=black, line width=0.3pt},
        server group/.pic={
            \foreach \i in {1,...,5} {
                \node[server] (-serv\i) at (\i*4.5pt-12pt, -12pt) {};
            }
        }
    ]

        \def\max{4}

        \begin{pgfonlayer}{background}
            \foreach \x in {0,...,\max} {
                \foreach \y in {0,...,\max} {
                    \pgfmathtruncatemacro\nextx{\x+1}
                    \ifnum\nextx<5
                    \foreach \targetx in {\nextx,...,\max} {
                        \draw[cable] (\x,\y) edge[out=45, in=135, looseness=0.4] (\targetx,\y);
                    }
                    \fi
                    \pgfmathtruncatemacro\nexty{\y+1}
                    \ifnum\nexty<5
                    \foreach \targety in {\nexty,...,\max} {
                        \draw[cable] (\x,\y) edge[out=-45, in=45, looseness=0.4] (\x,\targety);
                    }
                    \fi
                }
            }
        \end{pgfonlayer}

        \foreach \x in {0,...,\max} {
            \foreach \y in {0,...,\max} {
                \node[switch] (sw\x\y) at (\x, \y) {\x,\y};

                \begin{pgfonlayer}{background}
                    \pic[rotate=-20] (sg\x\y) at (sw\x\y) {server group};
                \end{pgfonlayer}

                \foreach \i in {1,...,5} {
                    \draw[serverlink] (sw\x\y) -- (sg\x\y-serv\i);
                }
            }
        }

    \end{tikzpicture}
    \caption{HyperX $5 \times 5$ topology with 5 servers per switch.}
    \label{fig:hx_5x5}
\end{figure}
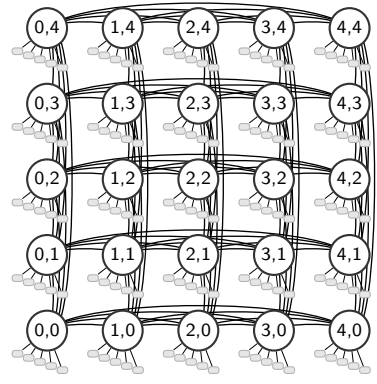

	Figure~\ref{fig:hx_5x5} shows a 2D HyperX with 5 switches per dimension and 125 endpoints.
	Observe that every row and column is a complete graph and, consequently, the network diameter is 2.

	\subsection{Routing}

To deal with different traffic patterns, the network must be equipped with an adaptive routing mechanism. A well-established routing for HyperX topologies is Omni-WAR~\cite{Kim_omni}. A closely related mechanism that uses the same set of routes is DAL~\cite{HyperX}.

	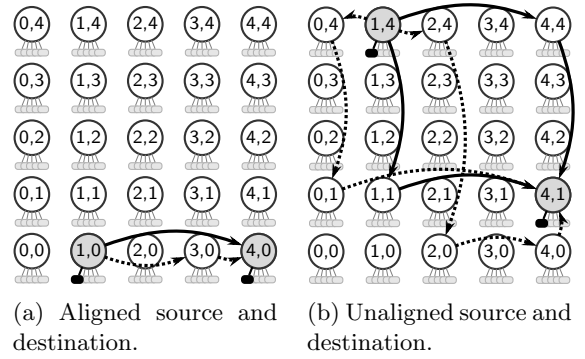
\begin{figure}
    \centering%
	\tikzset{
		arrow/.tip={Stealth[length=2.0mm, width=1.2mm]},
	}
	\hfill
    \begin{subfigure}[t]{0.45\linewidth}
        \centering
        \usetikzlibrary{backgrounds, arrows.meta, calc}
        \pgfdeclarelayer{background}
        \pgfsetlayers{background,main}
        \begin{tikzpicture}[
            scale=0.5,
            x=1.5cm, y=1.5cm,
            switch/.style={
                draw=black!80, circle, thick, fill=white,
                inner sep=0pt, outer sep=-0.1pt, minimum width=13pt,
                font=\scriptsize\sffamily
            },
            server/.style={
                draw=gray!60, rectangle, rounded corners=1pt, fill=gray!20,
                inner sep=0pt, minimum width=4pt, minimum height=2.5pt
            },
            highlighted_server/.style={
                draw=black, thick, fill=black,
            },
            cable/.style={draw=black!30, line width=0.5pt},
            serverlink/.style={draw=black!50, line width=0.3pt},
            highlighted_serverlink/.style={
                draw=black, line width=0.8pt,
            },
            min_path/.style={overlay,draw=black, line width=1.2pt, -arrow},
            non_min_path/.style={overlay,draw=black, line width=1.2pt, densely dotted, -arrow},
            h_inc/.style={bend left=20},
            h_dec/.style={bend right=20},
            v_inc/.style={bend right=20},
            v_dec/.style={bend left=20}
        ]

            \def\max{4}

            \foreach \x in {0,...,\max} {
                \foreach \y in {0,...,\max} {
                    \node[switch] (sw\x\y) at (\x, \y) {\x,\y};
                    \begin{scope}[shift={($(sw\x\y.south) + (0,-0.1)$)}]
                    \foreach \i in {1,...,5} {
                        \begin{pgfonlayer}{background}
                            \node[server] (serv\x\y\i) at (\i*0.1-0.3, -0.1) {};
                        \end{pgfonlayer}
                        \draw[serverlink] (sw\x\y) -- (serv\x\y\i);
                    }
                    \end{scope}
                }
            }

            \node[switch, fill=gray!30] at (1,0) {1,0};
            \node[switch, fill=gray!30] at (4,0) {4,0};

            \node[server, highlighted_server] (serv10h) at ($(sw10.south) + (-0.2, -0.2)$) {};
            \draw[highlighted_serverlink] (serv10h) -- (sw10);

            \node[server, highlighted_server] (serv40h) at ($(sw40.south) + (-0.2, -0.2)$) {};
            \draw[highlighted_serverlink] (sw40) -- (serv40h);

            \draw[min_path] (sw10) edge[h_inc] (sw40);

            \draw[non_min_path] (sw10) edge[h_dec] (sw30);
            \draw[non_min_path] (sw30) edge[h_dec] (sw40);

        \end{tikzpicture}
        \caption{Aligned source and destination.}
        \label{fig:omni_routing_5x5_a}
    \end{subfigure}
    \hfill
    \begin{subfigure}[t]{0.45\linewidth}
        \centering
        \usetikzlibrary{backgrounds, arrows.meta, calc}
        \pgfdeclarelayer{background}
        \pgfsetlayers{background,main}
        \begin{tikzpicture}[
            scale=0.5,
            x=1.5cm, y=1.5cm,
            switch/.style={
                draw=black!80, circle, thick, fill=white,
                inner sep=0pt, outer sep=-0.1pt, minimum width=13pt,
                font=\scriptsize\sffamily
            },
            server/.style={
                draw=gray!60, rectangle, rounded corners=1pt, fill=gray!20,
                inner sep=0pt, minimum width=4pt, minimum height=2.5pt
            },
            highlighted_server/.style={
                draw=black, thick, fill=black,
            },
            cable/.style={draw=black!30, line width=0.5pt},
            serverlink/.style={draw=black!50, line width=0.3pt},
            highlighted_serverlink/.style={
                draw=black, line width=0.8pt,
            },
            min_path_xy/.style={overlay,draw=black, line width=1.2pt, -arrow},
            min_path_yx/.style={overlay,draw=black, line width=1.2pt, dashed, -arrow},
            non_min_path/.style={overlay,draw=black, line width=1.2pt, densely dotted, -arrow},
            h_inc/.style={bend left=20},
            h_dec/.style={bend right=20},
            v_inc/.style={bend right=20},
            v_dec/.style={bend left=20}
        ]

            \def\max{4}

            \foreach \x in {0,...,\max} {
                \foreach \y in {0,...,\max} {
                    \node[switch] (sw\x\y) at (\x, \y) {\x,\y};
                    \begin{scope}[shift={($(sw\x\y.south) + (0,-0.1)$)}]
                    \foreach \i in {1,...,5} {
                        \begin{pgfonlayer}{background}
                            \node[server] (serv\x\y\i) at (\i*0.1-0.3, -0.1) {};
                        \end{pgfonlayer}
                        \draw[serverlink] (sw\x\y) -- (serv\x\y\i);
                    }
                    \end{scope}
                }
            }

            \node[switch, fill=gray!30] at (1,4) {1,4};
            \node[switch, fill=gray!30] at (4,1) {4,1};

            \node[server, highlighted_server] (serv14h) at ($(sw14.south) + (-0.2, -0.2)$) {};
            \draw[highlighted_serverlink] (serv14h) -- (sw14);

            \node[server, highlighted_server] (serv41h) at ($(sw41.south) + (-0.2, -0.2)$) {};
            \draw[highlighted_serverlink] (sw41) -- (serv41h);

            \draw[min_path_xy] (sw14) edge[h_inc] (sw44);
            \draw[min_path_xy] (sw44) edge[v_dec] (sw41);
            \draw[min_path_xy] (sw14) edge[v_dec] (sw11);
            \draw[min_path_xy] (sw11) edge[h_inc] (sw41);

            \draw[non_min_path] (sw14) edge[h_dec] (sw04);
            \draw[non_min_path] (sw04) edge[v_dec] (sw01);
            \draw[non_min_path] (sw01) edge[h_inc] (sw41);

            \draw[non_min_path] (sw14) edge[h_dec] (sw24);
            \draw[non_min_path] (sw24) edge[v_dec] (sw20);
            \draw[non_min_path] (sw20) edge[h_inc] (sw40);
            \draw[non_min_path] (sw40) edge[v_inc] (sw41);

        \end{tikzpicture}
        \caption{Unaligned source and destination.}
        \label{fig:omni_routing_5x5_b}
    \end{subfigure}
	\hfill
    \caption{Minimal (solid) and non-minimal routing paths (dotted) in Omni-WAR routing in a $5 \times 5$ HyperX. Links are omitted for clarity.}
    \label{fig:omni_routing_5x5}
\end{figure}

	The algorithm for Omni-WAR routing works over the difference in coordinates from the current switch to the destination switch.
	From the source switch, a dimension is said to be \textsl{unaligned} if it has a different value than the destination switch.
	Omni-WAR proceeds hop by hop, allowing a packet to move only through unaligned dimensions.
	Inside an unaligned dimension there is a single minimal hop and many potential deroutes.
	The minimal hop is given a preference over the deroutes, which is combined with their occupancies to obtain the selected port.

	The routing employs a hop-count mechanism to avoid both livelocks and deadlocks, a limit of $m$ non-minimal hops is imposed, forcing hops beyond this limit to be minimal.
	This is a global limit, specifically allowing more than one deroute in the same dimension.
	Thus, it allows each packet to perform up to $q + m$ hops, where $m$ is the non-minimal hops permitted, and $q$ is the number of dimensions in the HyperX. For some adverse traffic patterns it is necessary that $m\geq q$ to achieve acceptable performance, and $m=q$ is always sufficient; this is assumed henceforth.

	The possible routes are illustrated in Figure~\ref{fig:omni_routing_5x5}. For aligned switches, (a), there is a single minimal path and all switches visited by any route belong to the same row. For unaligned switches, (b), there are two minimal paths and any switch can be visited by some route.

	\subsection{Scalability}

	\begin{figure}
		\centering
		\includegraphics[scale=0.5]{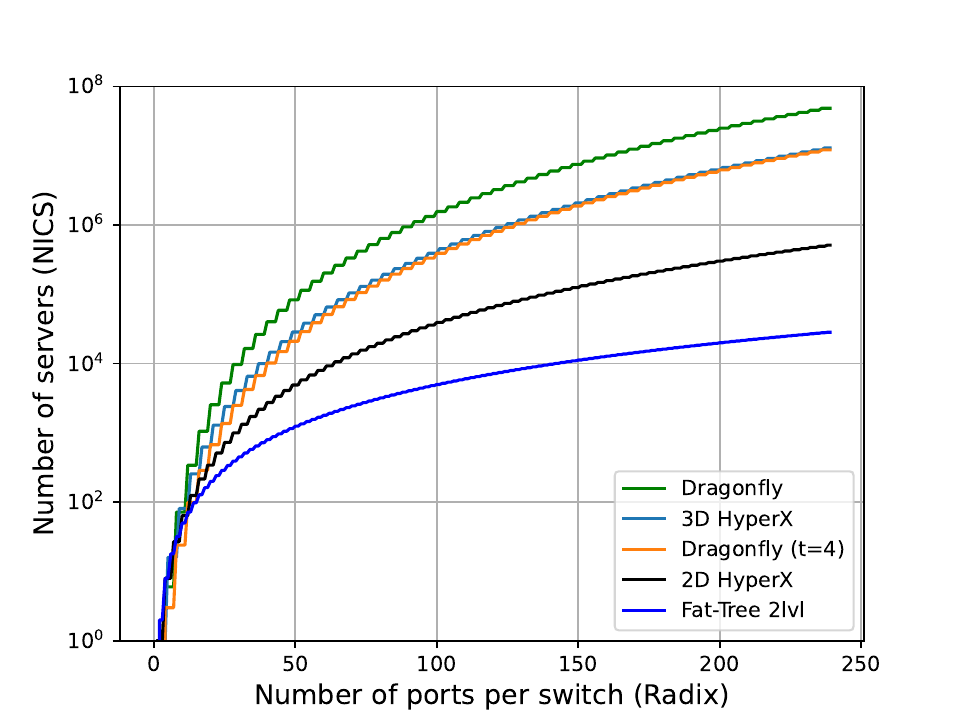}%
		\caption{Scalability of HyperX and other topologies.}
		\label{fig:scalabilty}
	\end{figure}

	A key advantage of HyperX is its ability to offer excellent scalability at a relatively low cost.
	Figure~\ref{fig:scalabilty} compares the scalability of HyperX with other network topologies, such as Fat-tree and Dragonfly, represented in Figure~\ref{fig:topologias}.
	The three top lines of the figure correspond to topologies with diameter 3 and the other two lines, at the bottom, to topologies of diameter 2. Note that the number of cables per endpoint grows proportionally to the diameter.

	While 3D HyperX is slightly less scalable than Dragonfly, the latter typically requires \textit{trunking} to be implementable in practice, which means that there are multiple links between groups. For example, in the Frontier supercomputer, its Dragonfly network is implemented with a trunking factor of $t = 4$, that is, there are 4 links between any pair of groups. It employs \prettynumber{2464} switches with radix 64 to provide \prettynumber{39424} endpoints connecting Network Interface Cards (NICs) from the server's GPUs.
	A 3D HyperX of dimensions $16 \times 16 \times 16$ using switches with the same radix, would employ \prettynumber{4096} switches to provide \prettynumber{65536} endpoints.
	Thus, the scalability of a 3D HyperX is a bit above a Dragonfly with its required trunking.

	As examples of diameter 2, consider using again switches with radix 64. A two-level Fat-tree can support \prettynumber{2048} endpoints whereas a 2D HyperX with $22 \times 22$ switches provides \prettynumber{10648} endpoints.
	Using switches with radix 128, a two-level Fat-tree can support \prettynumber{8192} endpoints whereas a 2D HyperX with $43 \times 43$ switches provides \prettynumber{79507} endpoints.
	Hence, among diameter-two topologies, 2D HyperX clearly outperforms leaf-spine Fat-trees in terms of scalability.

	Given this balance of cost, scalability, and simplicity, this work focuses on the 2D HyperX topology as a strong candidate for future HPC systems.
	Moreover, the findings presented in this study can be naturally extended to 3D HyperX configurations.

\section{Preliminaries on resource allocation}\label{sec:background}

The communication pattern of a parallel application can be modeled as a graph, called the \textsl{application graph}, in which vertices represent the application’s ranks or processes, and edges connect pairs of ranks that communicate.
A vertex in the application graph is considered to occupy an entire endpoint. This could represent a MPI rank that uses multithreading (e.g., via OpenMP) to exploit all available cores on that endpoint. If modeling a pure MPI application with multiple ranks per endpoint, each application vertex should instead represent an aggregate of those ranks. Thus, an application vertex abstracts the communication generated and consumed by a single endpoint.

In a parallel machine, the application graph must be mapped onto physical endpoints. First, the \textsl{resource allocation function} selects a set of system endpoints to host the application, forming the \textsl{partition}. The process of mapping the application graph onto this partition is referred to as \textsl{task mapping}. As the term partition suggests, different applications are assigned disjoint partitions; that is, practices such as co-location are not considered here.

\begin{figure}[tbp]
	\centering%
	\begin{tikzpicture}[
		x=.5cm,y=.5cm,
		font=\scriptsize,
		circled subgraph/.style={blue,fill=blue!20,rounded corners=4pt,thick,fill opacity=0.4},
		process/.style={draw,rectangle,inner sep=0pt,minimum width=60pt,minimum height=5pt},
		server/.style={
			draw=gray!60, rectangle, rounded corners=1pt, fill=gray!20,
			inner sep=0pt, minimum width=60pt, minimum height=5pt
		},
		switch/.style={
			draw=black!80, circle, thick, fill=white,
			inner sep=0pt, minimum width=2pt,
			font=\scriptsize\sffamily
		},
	]
	\pgfmathtruncatemacro\lado{8}
	\pgfmathtruncatemacro\last{\lado - 1}
	\pgfmathtruncatemacro\plast{\lado - 2}
	\def\marg{.4}
	\begin{scope}
		\newlength\auxlength
		\setlength\auxlength{1ex}
		\foreach \y in {0,...,\last}
		{
			\node[process] (process \y) at (0,\y) {};
			\pgfmathtruncatemacro\firstprocess{8*\y}
			\pgfmathtruncatemacro\lastprocess{8*\y+7}
			\path (process \y.east) node[font=\tiny,yshift=-2pt,anchor=base west,inner sep=.5pt] (label process \y) {\makebox[\auxlength][r]{\firstprocess}--\makebox[\auxlength][r]{\lastprocess}};
		}
		\def\patw{30}
		\newlength\commline
		\setlength\commline{2pt}
		\foreach \pa/\pb in {0/1,2/3,4/5,6/7}
		{
			\pgfmathsetmacro\sft{\patw+5}
			\draw[line width=4*\commline] ([xshift=\sft pt]process \pa.north west) -- ([xshift=\sft pt]process \pb.south west);
		}
		\foreach \pa/\pb in {0/2,1/3,4/6,5/7}
		{
			\pgfmathsetmacro\sft{\patw+11.5+Mod(\pa,2)*4.5}
			\draw[line width=2*\commline] ([xshift=\sft pt]process \pa.north west) -- ([xshift=\sft pt]process \pb.south west);
		}
		\foreach \pa/\pb in {0/4,1/5,2/6,3/7}
		{
			\pgfmathsetmacro\sft{\patw+19.5+\pa*2.5}
			\draw[line width=\commline] ([xshift=\sft pt]process \pa.north west) -- ([xshift=\sft pt]process \pb.south west);
		}
		\foreach \y in {0,...,\last}
		{
			\path[pattern=crosshatch] (process \y.north west) -- (process \y.south west) -- ++(\patw pt,0) -- ++(0,5pt) --cycle;
		}
		\path (label process 7.north east) ++(3pt,1.5pt) node[anchor=base east] {process ranks};
	\end{scope}
	\begin{scope}[xshift=2cm]
		\foreach \x in {0,...,\last}
		{
			\foreach \y in {0,...,\last}
			{
				\path (\x,\y) node[switch] (HX \x \y) {};
			}
		}
		\foreach \x in {0,...,\plast}
		\foreach \y in {0,...,\last}
		{
			\pgfmathtruncatemacro\nx{\x+1}
			\foreach \ox in {\nx,...,\last}
			{
				\draw[opacity=0.1] (HX \x\y) edge[overlay,line cap=round,looseness=0.4,out=45,in=90+45] (HX \ox\y);
			}
		}
		\foreach \x in {0,...,\last}
		\foreach \y in {0,...,\plast}
		{
			\pgfmathtruncatemacro\ny{\y+1}
			\foreach \oy in {\ny,...,\last}
			{
				\draw[opacity=0.1] (HX \x\y) edge[overlay,line cap=round,looseness=0.4,out=45,in=-45] (HX \x\oy);
			}
		}
		\foreach \k in {0}
		{
			\pgfmathtruncatemacro\tmp{Mod(\k,\lado)}
			\edef\drawcolor{red}
			\draw[circled subgraph,draw=\drawcolor,fill=\drawcolor!20] (\k,-\marg) -- ++(\lado-1+\marg,\lado-1+\marg) -- ++(-\marg,\marg) -- ++(-\lado+1-\marg,-\lado+1-\marg) --cycle;
		}
		\node[anchor=north] at (.5*\lado,0) {switches in a HX};
	\end{scope}
	\begin{scope}[yshift=-4cm]
		\foreach \y in {0,...,\last}
		{
			\node[process] (zoom process \y) at (0,\y) {};
			\path (zoom process \y.east) node[font=\tiny,yshift=-2pt,anchor=base west,inner sep=.5pt] (label zoom process \y) {\y};
		}
		\setlength\commline{2pt}
		\foreach \pa/\pb in {0/1,2/3,4/5,6/7}
		{
			\pgfmathsetmacro\sft{5}
			\draw[line width=4*\commline] ([xshift=\sft pt]zoom process \pa.north west) -- ([xshift=\sft pt]zoom process \pb.south west);
		}
		\foreach \pa/\pb in {0/2,1/3,4/6,5/7}
		{
			\pgfmathsetmacro\sft{11.5+Mod(\pa,2)*4.5}
			\draw[line width=2*\commline] ([xshift=\sft pt]zoom process \pa.north west) -- ([xshift=\sft pt]zoom process \pb.south west);
		}
		\foreach \pa/\pb in {0/4,1/5,2/6,3/7}
		{
			\pgfmathsetmacro\sft{19.5+\pa*2.5}
			\draw[line width=\commline] ([xshift=\sft pt]zoom process \pa.north west) -- ([xshift=\sft pt]zoom process \pb.south west);
		}
	\end{scope}
	\path (label zoom process 7.east) ++(2,-.5) node[draw,thick,circle,inner sep=0pt,minimum width=10pt] (zoom switch) {};
	\path (zoom switch.east) ++(1ex,0) node[anchor=west] {zoom of switch 0,0};
	\foreach \y in {0,...,\last}
	{
		\path (zoom switch.south) ++(0,-80pt+\y*10pt) node[server,anchor=west] (zoom server \y) {};
	}
	\foreach \y in {1,...,\last}
	{
		\draw (zoom switch) -- ++(5*\y:10pt);
		\draw (zoom switch) -- ++(80+5*\y:10pt);
	}
	\foreach \y in {0,...,\last}
	{
		\draw (zoom switch) edge[looseness=0.9,out=-70+5*\y] (zoom server \y);
	}
	\draw[decorate,decoration={brace,mirror,raise=3pt}] (zoom server 0.south east) -- (zoom server 7.north east) node[midway,anchor=west,xshift=3pt]{servers};
	\path ($(HX 00)!.5!(zoom switch)$) coordinate (zoom middle);
	\path[name path=middle circle]
		let \p1=($(zoom middle)-(zoom switch)$), \n1={veclen(\x1,\y1)} in
		(zoom middle) circle (\n1);%
	\path[name path=zoom subcircle] (zoom switch) circle (4pt);%
	\path[name intersections={of=middle circle and zoom subcircle}]
		(intersection-1) circle (2pt) coordinate (int1)
		(intersection-2) circle (2pt) coordinate (int2);
	\foreach \int in {int1,int2}
	{
		\pgfmathanglebetweenpoints{\pgfpointanchor{zoom switch}{center}}{\pgfpointanchor{\int}{center}}
		\pgfmathsetmacro{\angle}{\pgfmathresult}
		\draw[help lines] (HX 00.\angle) -- (zoom switch.\angle);
	}
	\foreach \p in {0,...,\last}
	{
		\draw[{Stealth[black,length=3pt,round,reversed]}-{Stealth[black,width=3pt,round]}] (label process \p) edge[,line width=.1pt,looseness=0.6,draw=white,double=black,double distance=.6pt,out=20] (HX \p \p);
	}
	\foreach \p in {0,...,\last}
	{
		\draw[{Stealth[black,length=3pt,round,reversed]}-{Stealth[black,width=3pt,round]}] (label zoom process \p) edge[,line width=.1pt,looseness=0.6,draw=white,double=black,double distance=.6pt,out=20,in=180] (zoom server \p.west);
	}
	\end{tikzpicture}%
	\caption{Inter-group communication from reduce scatter a la Rabenseifner of 64 processes with 8 groups of 8 processes each. Each group is mapped to the servers attached to one switch of a $8\times 8$ HX. The 8 selected switches lie in a diagonal of the topology.}\label{fig:dibujo_mapeo}
\end{figure}
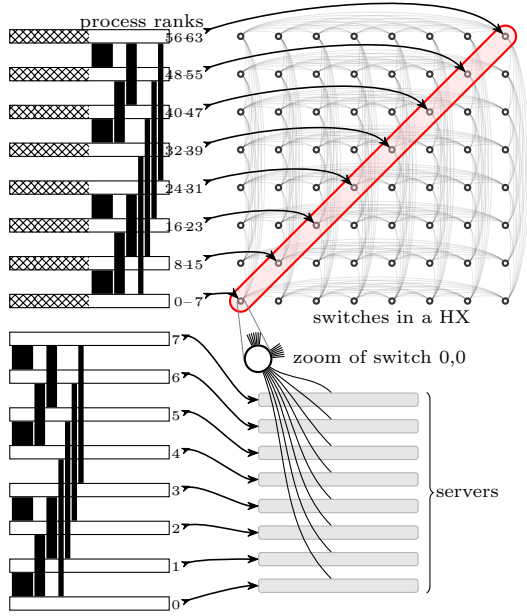

From a graph-theoretical perspective, the application graph is the \textsl{guest graph}, which is embedded into the system topology or \textsl{host graph}. The composition of the resource allocation function and task mapping defines an embedding function that maps vertices of the guest graph onto vertices of the host graph. When considering multiple applications, the embedding can be considered jointly, by taking as guest graph the disjoint union of all applications graphs. Note that the host graph represents switches, but allocation units are endpoints. Thus, the graph embedding is many-to-one, limited to the number of endpoints per switch. This terminology follows prior work and the functions we provide are always defined on the vertices of multiple applications onto the set of network endpoints.
Figure~\ref{fig:dibujo_mapeo} illustrates the idea. The application graph can be inferred from the communication pattern shown on the left, where each process communicates with its neighbors in a hypercube graph. The host graph corresponds to a 2D HyperX topology, and under certain resource allocation function, the guest graph is mapped onto the set of endpoints belonging to the switches along the selected diagonal.

Different strategies for resource allocation and task mapping lead to distinct challenges and trade-offs. One particularly relevant concept for understanding the impact of these strategies is the \textit{dilation}, which is already considered in \cite{Aleliunas, Moreira} and is adapted here for convenience in the following definition.

\begin{definition}\label{def:dilation}
	Let $f$ be a graph embedding of the graph $G(V, E)$ onto the graph $G'(V', E')$. Then, the \textsl{dilation} of the edge $\{A,B\}\in E$ is defined as the length of a shortest path between $f(A)$ and $f(B)$ in $G'(V', E')$.
\end{definition}

Therefore, key metrics for evaluating a resource allocation strategy include the average and maximum distances induced by the allocation function $f$ within the selected system partition, or host graph. These metrics directly bound the dilation of any application graph mapped onto the host graph.

Another important concept is convexity, which indicates whether an application is isolated, as it controls the interference between sets of switches. This property is formalized by the following definition.

\begin{definition}\label{def:convexity}
	Let $G=(V, E)$ be a graph. A set $S \subseteq V$ is \textsl{convex} if for every pair of vertices in $S$, every minimal path joining them lies entirely within $S$. The set $S$ is weakly-convex \cite{Anand} if for every pair of vertices, at least one minimal path lies entirely within $S$.
\end{definition}

We introduce next another property that indicates whether the set of endpoints of a switch is associated to a single partition.
\begin{definition}\label{def:switch-locality}
	A partition $\mathcal P$ is said to have \textsl{switch locality} if for every switch $x$ connected to an endpoint in $\mathcal P$, every endpoint connected to $x$ is also in $\mathcal P$.
	A resource allocation function that provides partitions with switch locality is called \textsl{locality-aware}.
\end{definition}

The problem of resource allocation has been widely considered in well-known topologies, such as Fat-tree, Torus and Dragonfly networks. Next, main findings are summarized to motivate the present study on HyperX resource allocation.

\subsection{Fat-Tree}

Fat-tree topologies are widely used in HPC systems due to their high bisection bandwidth and path diversity, as demonstrated in large-scale systems such as Eagle and Summit~\cite{TOP500}.

Fat-trees are non-blocking when fully utilized, allowing for flexible resource allocation without significant performance degradation.
However, resource allocation in Fat-trees typically focuses on packing jobs into subtrees to minimize interference between applications~\cite{jokanovic2015quiet}.

A tree is a convex graph. A pod is technically just weakly-convex, as its top-level switches introduce indirect paths through the root layer.
This is a consequence of applying graph-theoretic convexity to an indirect network and is not central to our discussion.
When endpoints within the pod communicate, no link outside the pod is utilized.
Thus, convexity becomes a key consideration in Fat-tree allocations, as allocating jobs within convex subtrees helps isolate traffic and limit congestion propagation.

\subsection{Torus}

Torus networks have been widely used in systems such as Blue Gene, Fugaku, and Google TPU architectures~\cite{Moreira,matsuoka2021fugaku,jouppi2023tpu}.
Formally, a torus is defined as the Cartesian graph product of cycles.
Figure~\ref{fig:topologias} illustrates an example of a 3D torus.

Torus topologies are particularly well-suited for applications that exhibit strong spatial locality. In such cases, the multi-dimensional structure of the torus helps to preserve communication locality and reduce contention.
However, a significant challenge with torus networks is the potential for long communication paths.
This makes dilation a particularly critical metric in Torus based systems, as increased dilation can directly lead to performance degradation.

There are two main strategies for resource allocation in Tori: partitioning, which divides the network into mesh-like regions using regular cuts~\cite{particionestoros}, and reconfiguration, which enables more flexible, disjoint partitions by dynamically reshaping the topology. The latter includes systems such as the IBM Blue Gene family and Google TPUv4. In these systems, resource-allocation consists of reconnecting, via link-chips or optical switches, basic blocks into a torus dedicated to the application~\cite{Gara,jouppi2023tpu}. These basic blocks are meshes that span one or half a rack; some instances of used partitions can comprise up to 64 such blocks wired into a single Torus. Task mapping is approached geometrically as in the strategies based on partitioning, under the assumption that the whole Torus is allocated to the application~\cite{Moreira}.
Other parallel machines, such as the Cray XK series, employ a Torus interconnect that is not reconfigurable, limiting resource allocation to fixed partitions. In both cases, the goal is to minimize dilation by allocating convex regions.

\subsection{Dragonfly}

High-radix, low-diameter interconnects—such as \textsl{Dragonflies}—are being increasingly used in modern HPC systems. Resource allocation in these high-radix topologies differs fundamentally from that of traditional networks.

First, they are low-diameter networks, where the maximum distance between any pair of switches is typically limited to $2$ or $3$ hops, values that bound the dilation.
Second, performance is primarily constrained by bisection bandwidth. A fundamental characteristic of these topologies is that, while achieving optimal throughput under uniform random traffic, they generally do not provide full bisection bandwidth, making the network susceptible to severe degradation under adversarial traffic. To mitigate this, adaptive, non-minimal routing is essential to redistribute load across all available links. Thus, resource allocation schemes that prioritize maximizing both minimal and non-minimal path diversity become crucial.

For instance, a representative allocation policy for Dragonfly networks is \textsl{LevelSpread}~\cite{level-spread}. In a full-sized Dragonfly, any pair of groups is connected by a single global link, which constitutes a bottleneck under heavy load. A poorly placed job can easily saturate these global links, severely impacting network performance. Allocating a job into just two groups causes half the endpoints of the job to be connected primarily through a single global link. Unlike conventional schemes, LevelSpread does not focus on minimizing dilation or preserving convexity. Instead, it maximizes available bandwidth by distributing tasks across multiple groups in a round-robin fashion when a job exceeds the capacity of a single group. However, when a job exactly fits within a single group, LevelSpread prioritizes compact placement inside the group, as the bisection bandwidth is high within a group, and this behaves similarly to a rearrangeably non-blocking subnetwork—assuming no external traffic is routed through it.

Overall, allocation objectives in Dragonflies shift from minimizing dilation and enforcing convexity toward maximizing path diversity and bandwidth.

	\section{A Proposal of Allocation Functions for HyperX Networks}
	\label{sec:resource_selection}

	Although both Dragonfly and HyperX are high-radix, low-diameter networks, their internal organization differs significantly.
	While Dragonfly has high internal bandwidth inside each group and limited global bandwidth between groups, HyperX uses a symmetrical structure with all-to-all connections without network bottlenecks and higher path diversity. Consequently, allocation strategies like LevelSpread, designed for Dragonfly, do not apply to HyperX.
	Instead, HyperX opens new possibilities for resource allocation, shifting the focus toward exploiting its symmetry and connectivity.
	This section presents the foundations of them.

As stated before, we concentrate on the 2D HyperX topology with diameter 2.
	In a system based on this network, an application can request a partition of any size, but to simplify the present study we focus on partitions of size $n^2$.
	Thus, we assume a scenario where a partition $\mathcal P$ is defined to contain exactly $n^2$ endpoints.
	A representative example of such a partition is a set of $n$ switches, where each switch is connected to $n$ different endpoints (a concentration factor of $n$).

\begin{figure}[tbp]
	\centering%
	\begin{tikzpicture}[
		x=.36cm,y=.36cm,%
		circled subgraph/.style={blue,fill=blue!20,rounded corners=3pt,thick,fill opacity=0.4},
		pin distance=5ex,%
		nothing/.style={inner sep=0pt},
		label/.style={semithick,draw,color=black,fill=white,fill opacity=0.9,font=\large, text width=1.8cm},
		every pin edge/.style={semithick},
	]
	\def\regioncolors{red,green,brown,blue,olive,gray,violet,cyan}
	\foreach \color [count=\i from 0] in \regioncolors
	{
		\expandafter\xdef\csname regioncolor-\i\endcsname{\color}
	}
	\def\marg{.4}
	\pgfmathtruncatemacro\lado{8}
	\pgfmathtruncatemacro\last{\lado - 1}

	\begin{scope}[xshift=0cm]
		\foreach \x in {0,...,\last}
		{
			\foreach \y in {0,...,\last}
			{
				\draw (\x,\y) circle (1pt);
			}
		}
		\foreach \k in {0,...,\last}
		{
			\edef\drawcolor{\csname regioncolor-\k\endcsname}
			\draw[circled subgraph,draw=\drawcolor,fill=\drawcolor!20] (-\marg,\k-\marg) rectangle ++(\lado-1+2*\marg,+2*\marg);%
		}
		\coordinate (origin row) at (0,0);
		\node[anchor=north] at (.5*\lado-.5,0) {\strut rows};
	\end{scope}
	\begin{scope}[xshift=3cm]
		\foreach \x in {0,...,\last}
		{
			\foreach \y in {0,...,\last}
			{
				\draw (\x,\y) circle (1pt);
			}
		}
		\path[clip] (-.5,-3) rectangle (\lado-.5,\lado-.5);
		\foreach \k in {-\last,...,\last}
		{
			\pgfmathtruncatemacro\tmp{Mod(\k,\lado)}
			\edef\drawcolor{\csname regioncolor-\tmp\endcsname}
			\draw[circled subgraph,draw=\drawcolor,fill=\drawcolor!20] (\k,-\marg) -- ++(\lado-1+\marg,\lado-1+\marg) -- ++(-\marg,\marg) -- ++(-\lado+1-\marg,-\lado+1-\marg) --cycle;
		}
		\coordinate (origin row) at (0,0);
		\node[anchor=north] at (.5*\lado-.5,0) {\strut diagonal};
	\end{scope}

	\begin{scope}[xshift=0cm,yshift=-3.2cm]
		\foreach \x in {0,...,\last}
		{
			\foreach \y in {0,...,\last}
			{
				\draw (\x,\y) circle (1pt);
			}
		}
		\pgfmathtruncatemacro\lastky{\lado/2-1}
		\foreach \kx in {0,1}
		\foreach \ky in {0,...,\lastky}
		{
			\pgfmathtruncatemacro\tmp{\kx+2*\ky}
			\edef\drawcolor{\csname regioncolor-\tmp\endcsname}
			\draw[circled subgraph,draw=\drawcolor,fill=\drawcolor!20] (\kx*\lado*.5-\marg,2*\ky-\marg) rectangle ++(\lado/2-1+2*\marg,1+2*\marg);%
		}
		\coordinate (origin subplane) at (0,0);
		\node[anchor=north] at (.5*\lado-.5,0) {\strut rectangles};
	\end{scope}

\begin{scope}[xshift=3cm,yshift=-3.2cm]
		\foreach \x in {0,...,\last}
		{
			\foreach \y in {0,...,\last}
			{
				\draw (\x,\y) circle (1pt);
			}
		}
		\coordinate (origin cross) at (0,0);
		\node[anchor=north] at (.5*\lado-.5,0) {\strut L-shapes};
		\path[clip] (-.5,-.5) rectangle (\lado-.5,\lado-.5);
		\foreach \kx in {-\lado,...,\lado}
		\foreach \ky in {-1,...,1}
		{
			\pgfmathtruncatemacro\tmp{Mod(\kx,\lado)}
			\edef\drawcolor{\csname regioncolor-\tmp\endcsname}
			\draw[circled subgraph,draw=\drawcolor,fill=\drawcolor!20] (\kx-\marg,\kx+\lado*\ky-\marg) -- ++(3+2*\marg,0) -- ++(0,2*\marg) -- ++(-3,0) -- ++(0,4) -- ++(-2*\marg,0) --cycle;
		}
	\end{scope}

	\end{tikzpicture}%
	\caption{Some linear and tile based partitions for resource allocation. Each colored region is a set of 8 switches and 64 servers where a job can be allocated.}\label{fig:formas_teselas}
\end{figure}
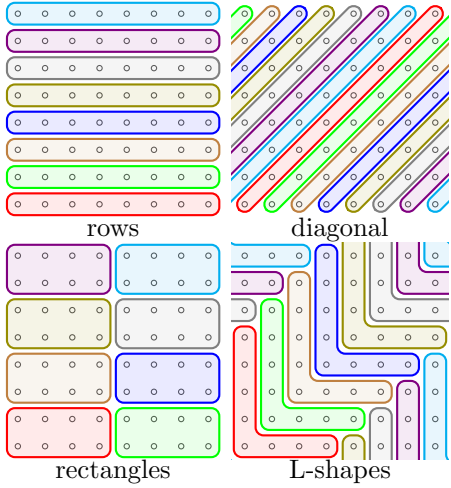

	In a 2D HyperX comprised of $n \times n$ switches with concentration $n$, the system capacity supports exactly $n$ non-overlapping partitions of size $n^2$.
	Each partition $\mathcal P_p$ is uniquely identified by $p\in \{0, \dots, n-1\}$.
	Examples of such partitions are represented in Figure~\ref{fig:formas_teselas} and will be later studied in detail.
	In practical implementations, these mappings are built incrementally.
	As applications are launched, the resource manager identifies a partition of idle nodes satisfying the requirements.
	Following this allocation, a secondary local task mapping function binds specific ranks from application to the endpoints within the allocated partition.

	We define the logical space inside of a partition $\mathcal P_p$ by labeling its endpoints with a linear index $r \in \{0, \dots, n^2-1\}$.
	To map this linear space onto the network topology, it is advantageous to decompose $r$ into a two-dimensional logical coordinate system using the Euclidean division:
	\begin{equation*}
		r = n \cdot r_y + r_x, \quad \text{where } 0 \leq r_x < n.
	\end{equation*}
	Here, $r_x = r \pmod n$ and $r_y = \lfloor r/n \rfloor$.

	Accordingly, the resource allocation policy is formalized as a mapping function $f$ that translates these logical coordinates into physical topology coordinates,

	\begin{equation}
		\label{eq:function-cube}
		f(p, r_y, r_x) = (s_y, s_x, c),
	\end{equation}

	\noindent where
	\begin{itemize}
		\item $p$ denotes the partition identifier.
		\item $(s_y, s_x)$ denotes the decomposed coordinates of the physical switch $s$, such that $s = s_y \cdot n + s_x$.
		\item $c$ denotes the local index of the endpoint connected to switch $s$ (offset within the switch).
	\end{itemize}

 Next, the proposed allocation policies are studied in
terms of the nature of the mapping functions: linear, geometric-based or stochastic.

	\subsection{Linear Resource Allocation Functions}

	A class of deterministic allocation functions that can be expressed as linear functions of the mapping $f(p, r_y, r_x) = (s_y, s_x, c)$ are considered in this subsection.

	The \textit{Row Selection} function concentrates a partition's resources within a single dimension of the switch grid.
	This function ensures that all endpoints reside in switches sharing the same row, resulting in a maximum partition distance of 1 hop.
	Formally, this corresponds to the identity function defined as:
	\begin{equation*}
		\operatorname{row}(p, r_y, r_x) = (p, r_y, r_x).
	\end{equation*}

	In contrast, the \textit{Full Spread} function maximizes distribution by allocating exactly one endpoint from every switch in the system. This is achieved by mapping the partition identifier to the local endpoint offset, thereby exposing the partition to the global bisection bandwidth of the topology:
	\begin{equation*}
		\operatorname{full\_spread}(p, r_y, r_x) = (r_y, r_x, p).
	\end{equation*}

	Finally, the \textit{Diagonal Selection} function allocates resources along a topological diagonal.
	All inter-switch communication within the partition must traverse exactly 2 hops, representing the worst-case locality scenario for this topology:
	\begin{equation*}
		\operatorname{diagonal}(p, r_y, r_x) = (r_y, (r_y + p) \pmod n, r_x).
	\end{equation*}
	This resource allocation functions is the one employed in Figure~\ref{fig:dibujo_mapeo}.

	\subsection{Tiled Resource Allocation Functions}

	Since the set of nodes of a HyperX can be naturally set onto a grid, allocation functions based on geometric tessellations of the grid are also included. Given the Cartesian structure of the 2D HyperX, valid resource partitions can be constructed by tiling the switch plane with compact geometric shapes.
	Unlike the linear functions, these strategies focus on forming contiguous or semi-contiguous 2D structures.

	The \textit{Rectangular Tessellation} function partitions the network into compact rectangular blocks:
	\begin{multline*}
		\operatorname{rectangular}(p, r_y, r_x) = \\
		\big(\!\rem(r_y,2)+\frac{n}{2}\rem(p,2),
		\quo(r_y,2)+2\quo(p,2),
		r_x \big).
	\end{multline*}

	Alternatively, the \textit{L-shape Tessellation} function allocates resources to a set of switches forming an orthogonal L-shaped structure. This shape is constructed by uniting a vertical segment and a horizontal segment that intersect at a common vertex (the partition anchor). The mapping function is defined piecewise to distribute the logical coordinate $r_y$ across these two arms:
	\begin{multline*}
		\operatorname{L\_shape}(p, r_y, r_x) =\\
		\begin{cases}
			(p +r_y,p ,r_x)  \text{, for } r_y < \quo(n,2),\\
		(p, p + r_y-\quo(n,2)+1,r_x)  \text{, otherwise.}
		\end{cases}
	\end{multline*}
	Modular arithmetic (modulo $n$) is implicitly applied to the resulting switch coordinates.
	This configuration represents a compromise between dimension-ordered alignment and local clustering.

	\subsection{Randomized Resource Allocation Functions}

	Randomization techniques have been usually employed to balance the use of resources and avoid hotspots.
	Moreover, randomization may help to remove bias from artificial choices.
	Next, the resource allocation functions based on randomization are formalized.

	The \textit{Random Endpoint Selection} function performs a complete scattering of resources, ignoring topological locality entirely. Let $\pi$ be a random permutation of the set $\{0,\dotsc,n-1\}^3$, the set of the coordinates of the $n^3$ physical endpoints. That is, $\pi$ sends each triplet to a randomly chosen triplet. The allocation function maps the linear index of the requested resource directly to this permuted space:
	\begin{equation*}
		\operatorname{random\_endpoint}(p, r_y, r_x) = \pi(p,r_y,r_x)
	\end{equation*}
	Under this function, for any two endpoints from the same partition, the probability they reside in the same switch is roughly $1/n^2$ and of residing in the same dimension is about $2/n$. Consequently, the vast majority of intra-partition communication pairs are at a distance of 2 hops, with a small fraction at distance 1 or 0.

	Alternatively, the \textit{Random Switch Selection} function preserves switch locality while randomizing the selection of switches.
	Each application is placed on a partition of $n$ random switches and their $n^2$ total endpoints. By convention, let $r_y$ be used as selector of the switch and use $r_x$ to select the endpoint.
	Formally, let $\sigma$ be a random permutation of the set $\{0,\dotsc,n-1\}^2$ of coordinates all the $n^2$ switches. The allocation function is
	\begin{equation*}
		\begin{split}
			\operatorname{random\_switch}(p, r_y, r_x) &= (s_y, s_x, r_x), \\
			\text{where } (s_y, s_x) &= \sigma(p , r_y).
		\end{split}
	\end{equation*}
	In contrast to the previous case, groups of $n$ endpoints are at distance 0.
	However, the connectivity between these switch groups remains randomized; thus, inter-switch traffic predominantly traverses 2 hops, similar to the fully random approach.

	\section{Topological Properties of the Partitions}
	\label{sec:analysis}

The allocation functions proposed in the previous section are intended to cover most reasonable strategies.
In this section, the properties of the partitions resulting from these allocation functions are analyzed.
Three aspects are considered: distance, convexity, and a new metric introduced by the authors, referred to as \textsl{partition bandwidth}.

It should be noted that the HyperX topology admits many automorphisms \cite{Harary}, which makes some function formulations equivalent. The clearest example is the equivalence between Row and Column allocations. An equivalence also exists between L-shaped partitions and star-shaped partitions, that is, a central switch with four rays. Likewise, any diagonal partition is equivalent to any partition that selects exactly one switch in each row and each column.

	\subsection{Distance properties}

The average and maximum distance of a partition provide direct upper bounds on the average and maximum dilation of an application placed on that partition. Moreover, uniform applications attain these distances as their dilation. Distances are also fundamental for the analysis of other properties, such as partition bandwidth, which is studied later.

The impact of distances on performance is more limited in a low-diameter topology such as HyperX than in topologies such as a Torus. This does not imply that distance is unimportant, as it may contribute up to a factor of two in performance variation; rather, it is not the dominant factor. For the same reason, the bounds relating dilation and distance are necessarily tighter.

	The average distance of the endpoints of a partition $\mathcal P_p$, assuming that every source may have itself as destination, can be calculated as:

\begin{equation}
		D_{\mathcal P_p} = \frac{1}{|\mathcal P_p|}\sum_{e_1,e_2\in \mathcal P_p}d(e_1,e_2),
	\end{equation}
The term $d(e_1,e_2)$ denotes the number of switch-to-switch links in a minimal path, as before.

	As the simplest case, Full Spread selects one endpoint from each switch, which implies that the average and maximum distance of the partition coincide with the average and maximum distance of the topology. This is an average distance of $2-\frac{2}{n}$ and a diameter of 2 for the $n\times n$ HyperX.

In the case of locality-aware allocation functions, it is natural to consider the distance at the switch level, denoted as $D_S$, instead of the endpoint level. Let us find out that they match exactly.
Denote by $S$ the set of switches connected to the partition $\mathcal P_p$, which has switch locality.
Then, the sum of the distances between endpoints is $\sum_{e_1,e_2\in \mathcal P_p}d(e_1,e_2)=\sum_{s_1,s_2 \in S}n^2 d(s_1,s_2)$. Rewriting this expression as average distances results $|\mathcal P_p|^2D_{\mathcal P_p}=n^2|S|^2D_S$. Since $n|S|=|\mathcal P_p|$, it follows that $D_{\mathcal P_p}=D_S$.

	Consequently, it is indistinct to use endpoint versus switch terminology for these functions. Following the same type of calculations, the maximum and average distances can be obtained for the rest of the allocation functions, as follows:
	\begin{itemize}
		\item Row selects all the endpoints in a row, which means that all pairs of switches are at maximum distance of 1 and average distance of $1-1/n$.
		\item With Diagonal, no two selected switches share the same row or column. Then, the value for the maximum distance is 2 and average distance is $2-2/n$.
		\item In a Rectangular Tessellation the distance values are those of the smaller allowed HyperX subgraph.
			This is, a maximum distance of 2 and an average distance of $2-1/n_a-1/n_b$ for a $n_a\times n_b$ rectangle.
		\item In L-shape Tessellation, average distance is roughly $1+1/2$, as switches are divided into the two rays. Switches are at distance 1 when they are in the same ray and 2 when they are in different rays.
	\end{itemize}

Random allocation functions inherit their distance metrics from the host topology. Indeed, due to linearity of expectation, the distance between two random endpoints matches the average distance of Full Spread, which is the average distance of the whole HyperX topology. Similarly, the average switch distance of Random Switch and Random Endpoint coincides with the average switch distance of the whole HyperX topology. And by the above argument, for partitions with switch-locality it also coincides with the endpoint distance.

	\subsection{Convexity and locality}

	Convexity refers to whether the minimal paths between switches of a partition are contained inside it or not.
	If an application is mapped onto the endpoints of a convex set of switches, communications can be performed without affecting switches outside the set.
	This property can help to reduce interference, especially in well-established topologies such as Torus and Fat-tree, which typically employ minimal routing.
	However, in modern low-diameter topologies, such as Dragonfly and HyperX, even when using convex partitions, employing non-minimal routes that extend beyond the partition may be critical for performance.
	The significance of this depends on the traffic pattern, and its feasibility depends on the routing algorithm used.

	Row, Full Spread, and Rectangular Tessellation functions provide convex partitions.
	In Full Spread, the partition comprises all the network switches, so it is trivially convex.
	With Row, each horizontal link corresponds to the unique minimal path between their endpoints.
	With Rectangular Tessellation, the partitions are smaller HyperX instances within the whole network.

	The Diagonal function is the archetype of non-convexity since it enforces that there is no link between any switches of the partition.

	As an intermediate allocation function, L-shape Tessellation provides weakly-convex partitions.
	From the horizontal ray to the vertical ray the XY minimal path is contained, whereas the YX minimal path is not.
	In the opposite direction the ordering is reversed.

	Random functions provide non-convex partitions asymptotically almost surely except for the trivially convex cases comprising 0, 1 or $n^2$ switches, and the weak-convex case of $n^2-1$ switches. A Random Endpoint partition involves $n^2$ switches when it comprises more than $n^2\log n$ endpoints. For the ongoing assumption of $n^2$ endpoints, both random functions are non-convex.

Almost all the allocation functions are locality-aware, with the exception of the Random Endpoint Selection function and Full Spread, which are the ones that share switches among partitions.

	\subsection{Partition Bandwidth}
	\label{subsec:bandwidth_analysis}

	As it will be seen in Section \ref{sec:experiments}, neither distance nor convexity are topological properties of the partitions that enable a correct interpretation of the empirical results.
	For filling this gap, a new metric is introduced next. We define \textsl{partition bandwidth}, denoted as $\PB$, to the greatest rate at which load could be generated by every endpoint of a partition, with uniform destinations across such partition, and employing minimal routing. The $\PB$ metric gives an idea of how much total traffic could manage a partition before reaching its saturation point.
	To estimate the maximum bandwidth that a partition can support, computing $\PB$ assumes an idealistic scenario in which the endpoints of a partition can send uniform traffic across such partition at an unlimited rate.
	In other words, it indicates how well the partition is provisioned in total bandwidth. The $\PB$ metric leverages the following definition.

	\begin{definition} Given a graph $G=(V,E)$, the \textsl{convex hull} of a subset $S$ of the vertices is the subgraph induced by those edges in a shortest path between vertices in $S$.
	\end{definition}

	Only the links belonging to the convex hull of a partition can be employed to compute its partition bandwidth.
	Assuming all links are full-duplex, the maximum traffic load in the partition cannot exceed twice the number of links in its convex hull, denoted as $2L$.
	Since each packet injected by each endpoint of a partition traverses an average of $D_{\mathcal P_p}$ links, the following expression bounds its bandwidth.

	\begin{equation}\label{eq:throughput_symmetric}
		\PB\leq \frac{2L}{n^{2}D_{\mathcal P_p}}
	\end{equation}

	For symmetric cases, in which every link of the convex hull is used in the same proportion, it is obtained that this upper bound is actually an equality.
	This means that every endpoint within the partition could inject, at most, $\PB$ phits/cycle before saturating the partition.
	The $\PB$ values of the different partitions are considered next.

	Since the convex hull of a Row partition is a complete graph of $n$ switches, it has $\binom{n}{2}=\frac{n(n-1)}{2}$ links.
	This gives $\PB_{\textit{Row}}=\frac{2n(n-1)/2}{n^2\cdot (n-1)/n}=1$.

	In a Diagonal partition there are four times the links of the previous case, that is, $2n(n-1)$, and the average distance is the double, which results in $\PB_{\textit{Diagonal}}=2$.
	The convex hull of Full Spread is the whole topology with its $n^2(n-1)$ links, holding $\PB_{\textit{Full\,Spread}}=n$.

	For the Rectangle partition, let us consider the case in which one side doubles the other side for the calculation.
	In general, this is $\sqrt{2n}\times \sqrt{n/2}$, including the $4\times 2$ case used later in the experimentation.
	The bound (\ref{eq:throughput_symmetric}) gives $\PB_{\textit{Rectangle}}\leq \tfrac{3}{2\sqrt{2n}}$, but the available bandwidth is lesser since there are proportionally fewer links along the shortest dimension.
	Instead, the bound can be applied for each dimension, thus getting $\PB_{\textit{Rectangle}}=\min\{\PB_H,\PB_V\}=\PB_V$.
	The partition comprises $\sqrt{2n}\binom{\sqrt{n/2}}{2}\sim\tfrac{n^{1.5}}{2\sqrt{2}}$ vertical links, and the vertical average distance is $1-1/\sqrt{n/2}$.
	This results in $\PB_{\textit{Rectangle}}=\tfrac{1}{\sqrt{2n}}$.

	The convex hull of an L-shape partition comprises those links incident into the L-shape that are within the smallest rectangle that contains the L-shape.
	Consider the ray with $n/2$ switches beyond the center. Its $n^2/2$ endpoints, generating messages half the time in the same ray, are able to fit the $\binom{n/2}{2}$ links within. Approximately the same for the other ray. It can be seen that routing all other messages to the outside, also fit the outgoing links, giving $\PB_{\textit{L-shape}}\approx 1$. This asymptotically matches the value provided by (\ref{eq:throughput_symmetric}).

	With respect to random mappings, intuitively, Random Switch can be thought to be similar to Diagonal.
The chance of having aligned switches is small, giving an average distance close to 2 and using most links of each switch. However, the link utilization is not asymptotically 1. The number of rows covered by some switch is asymptotic to $n(1-e^{-1})$. This limits the number of available links, resulting in $\PB_{\textit{Random\,Switch}}\sim 2(1-e^{-1})\approx 1.26$.

Finally, Random Endpoint is similar to Full Spread. In this case, almost every link belongs to the convex hull, but again, not in an asymptotic sense. The number of switches in the partition is about $n^2(1-e^{-1})$, and for a link to be in the convex hull it is enough for any endpoint to belong to the partition. The chance for a link to belong to the convex hull is then $1-\left(1-\frac{n^2(1-e^{-1})}{n^2}\right)^2=1-e^{-2}$. The available bandwidth is then seen to be asymptotic to $\PB_{\textit{Random\,Endpoint}}\sim n(1-e^{-2})\approx 0.86n$.

	\begin{table*}
		\scriptsize
		\centering%
		\caption{Properties of the allocation functions and the partitions they provide. Considering partitions of $n^2$ endpoints. Rough round approximation for distance and partition bandwidth.}\label{tbl:analysis}%
		\begin{tabular}{|l|l|l|l|l|l|l|}
			\hline
			Allocation function			& Function type		& Avg. distance	& Max. distance 	& Convexity		& locality-aware & $\PB$ \\
			\hline
			Row            				& linear	& 1					& 1	& convex		& yes & 1\\
			Diagonal        			& linear	& 2					& 2	& non-convex	& yes & 2\\
			Full Spread					& linear	& 2					& 2	& convex		& no & $n$\\
			L-shape Tessellation		& tiling	& $1+1/2$			& 2	& weakly-convex	& yes & $1$\\
			Rectangular Tessellation	& tiling	& 2					& 2	& convex		& yes & $1/\sqrt{2n}$\\
			Random Endpoint        		& random	& 2					& 2	& non-convex	& no & $n(1-e^{-2})$\\
			Random Switch          		& random	& 2					& 2	& non-convex	& yes & $2(1-e^{-1})$\\
			\hline
		\end{tabular}
	\end{table*}

	\subsection{Partition's topological properties summary}

	The findings of the previous analysis are summarized in Table~\ref{tbl:analysis}.
	The content of the table, comparing the different partitions, reveals insights that will be contrasted against empirical results in Section \ref{sec:experiments}.

	With respect to distance properties, all partitions but Row and L-shape, share the same values.
	However, this up to $\times2$ improvement is overshadowed by bandwidth shortcomings.

	With respect to the convexity of partitions, there are three convex cases, other three non-convex and one more which is weakly-convex.
	Nevertheless, convexity loses some of its value when non-minimal routing has to be employed.
	This is the case of HyperX, in which partitions are naturally disturbed by remote traffic injected by other partitions.

	Larger differences can be observed with respect to partition bandwidth.
	Both Row partition and L-shape have $\PB_{\textit{Row}}=\PB_{\textit{L-shape}}=1$, Rectangular has $\PB_{\textit{Rectangular}}<1$ and the remainder partitions have $\PB>1$.

	This means that in Row, with a $\PB$ value of exactly 1, endpoints could inject uniform traffic at full rate.
	Remember that the actual network injection rate is 1 phit/cycle/endpoint.
	Injecting at this maximum rate means that there is no slack to absorb neither any external traffic nor any traffic non-uniformities in the partition, making the approach fragile.
	It could be good when all applications generate uniform traffic and fit nicely within a single row, provided that the routing is simple or guided enough to keep traffic confined to that row.
	It becomes limited when dealing with non-uniform traffic loads.

	Diagonal partition has $\PB_{\textit{Diagonal}}=2$, which ensures that, even injecting at full rate, the partition is able to manage twice as much traffic.
	This is an advantage for dealing with remote traffic in a fully utilized machine.
	In fact, it would be possible to handle adversarial traffic loads within a single partition at full rate by de-routing traffic through randomly selected switches across the partition.
	In spite of its non-convex nature, Diagonal partition is a strong candidate for maximizing global throughput.

	Full Spread, with $\PB_{\textit{Full\,Spread}}=n$, is a straightforward approach to ensure that all links in the network are utilized. However, it disregards switch locality and could result in complete interference.

	Random partitions have also $\PB>1$. Random Switch Selection is close to Diagonal in all parameters and Random Endpoint selection is close to Full Spread. Both have, however, a notable decrease in the number of links in their convex hull due to the expected coincidences. They can be a little worse, in general, than their deterministic counterparts.

	\section{Experimental Methodology}\label{sec:application_analysis}

In order to compare the proposed allocation techniques, an exhaustive experimental evaluation will be conducted.
The evaluation will consist of measuring the performance of different applications over a HyperX, in which the allocation strategy changes.
First, the set of employed workloads is described and analyzed.
Second, the experimental setup is detailed along with the two frameworks considered to evaluate the allocation functions.
Lastly, the simulated scenarios are outlined.

	\subsection{Simulated Workloads}

	Two kind of synthetic workloads are considered: static traffic patterns and kernels of real applications.
	The static traffic patterns are used to carry a preliminary analysis, while the kernels of real applications are used to validate the conclusions in a more realistic scenario.

	\subsubsection{Static traffic patterns}
		Messages used by these patterns are generated at a rate of 1 phit/cycle/endpoint, which is the maximum injection rate of the network.
The patterns considered are:
	\begin{itemize}
		\item \textbf{Uniform Random.} Each message is sent to a randomly selected destination within the same partition, with uniform (independent and identical) probability. It represents a benign scenario, where minimal routing is expected to perform well.
		\item \textbf{Random Permutation.} Each task sends messages to a unique destination, forming a random permutation of the tasks. This pattern is more challenging than the uniform random, as it creates a more structured load that can lead to congestion if not properly managed.
		\item \textbf{Random Switch Permutation.} Individual tasks are organized into groups of $n$ tasks, and between those groups a random permutation is applied. So, each group of tasks send messages only to another group of tasks. This pattern is extremely adversarial, as groups of tasks are mapped into unique switches when the allocation function preserves switch-locality and task mapping is linear. It requires the use of non-minimal routes to achieve good performance in most partitions.
	\end{itemize}

	\subsubsection{Communication kernels of real applications}
	They have been selected as representative kernels of communication patterns commonly found in HPC and AI workloads.
	The selection was guided by references from prior studies, and all chosen applications are characterized by intensive network usage.
	Next, they are detailed and their communication needs are discussed.

	The \textbf{All-to-All} MPI collective is a fundamental operation in which every process communicates with all other processes in the system.
	It is widely used in numerical applications such as the Fast Fourier Transform (FFT) and in AI applications based on model parallelism.
	When performed over $k$ processes, each rank performs a number of steps multiple of $k-1$. At step $i$, rank $r$ sends data to rank $r+i \mod k$, and receives data from rank $r-i \mod k$. Each rank sends and receives independently from the other ranks in a completely asynchronous manner.
	The operation completes once every task has exchanged messages with all other tasks, thereby distributing the network load evenly.
	Employing a larger amount of steps entails smaller chunks of data at every step, which makes the resulting communication more uniform. It is adjusted to make each chunk fit into a single packet.

	The \textbf{All-Reduce} operation is a collective communication pattern where all processes contribute data, and a reduction operation (e.g., sum, max, min) is applied to produce a single result distributed back to all processes.
	This pattern is critical in large-scale machine learning workloads.
	The algorithm used is the optimal Rabenseifner algorithm~\cite{Rabenseifner}.
	In this approach, processes are arranged in a hypercube, and the algorithm has two phases (Scatter-Reduce and All-Gather) of $\log k$ steps each.
	Variants of the algorithm for non-power-of-two process counts follow a similar pattern.

	\textbf{Stencil} patterns represent one-to-many communication, where each process interacts with its neighbors in a bidimensional space.
	The neighbor patterns used are \textsl{von Neumann Neighborhood}, in which communication occurs with immediate
	neighbors in the four cardinal directions, and \textsl{Moore Neighborhood}, where communication extends to diagonal neighbors.
	Stencil patterns are prevalent in scientific computing applications.
	Like the all-to-all operation, the communication can be organized in multiple rounds, allowing a more uniform and efficient use of the network.

	In the \textbf{Random Involution}, data is exchanged between randomly paired processes, as typically occurs in applications using unstructured or irregular grids, such as those arising in the finite element method.

	\subsection{Experimental Setup}

	One goal of the present study is to experimentally evaluate the resource allocation strategies described in the previous sections.
	This evaluation is done using the CAMINOS simulator~\cite{CAMINOS,CAMARERO2025105136}.
	CAMINOS is an event-driven network simulator that operates at flit granularity and models the switch microarchitecture.
	The parameters considered in each simulation are detailed in Table~\ref{tab:simulation_parameters}.

	\begin{table}
		\centering
		\caption{Simulation Parameters}
		\label{tab:simulation_parameters}
		\begin{tabular}{|l|l|}
			\hline
			\textbf{Parameter} & \textbf{Value} \\
			\hline
			Switch Internal Speedup & $2\times$ \\
			Allocator & Random \\
			Packet Size & 16 flits \\
			Input Buffer Size & 8 packets per VC\\
			Output Buffer Size & 4 packets per VC \\
			Number of VCs & 4 per port \\
			\hline
		\end{tabular}
	\end{table}

	The network evaluated is a 2D HyperX with $8 \times8$ switches, 8 endpoints per switch, giving a total of 512 endpoints.
	All the resource allocation strategies defined in Section~\ref{sec:resource_selection} are considered.
	To assess these strategies, the traffic patterns and application communication kernels described in previous subsection are used.
	The kernels are generated with three different sizes: 64, 128 and 256-process applications.
	For applications with 64 processes the partitions are exactly as described in Section~\ref{sec:resource_selection}. For applications with a number of processes multiple of 64, a partition consists on the union of consecutive blocks that would be partitions for 64. For example, Row Selection with 128 processes makes partitions of two consecutive rows.
	In the case of 256-process applications, the Row strategy allocates four complete rows, which coincides with the $8 \times 8 \times 4$ rectangle from the Rectangular Selection, and the later is omitted.

	The routing algorithm employed is Omni-WAR~\cite{Kim_omni}.
	However, minimal adaptive routing, MIN, is employed for a preliminary study. %
	Occupancy-based prioritization is performed, this is, the queue with least occupancy is selected, with deroutes being given an extra occupancy penalty, $P$, of 64 phits to discourage its use.
	Other routing algorithms, such as Polarized routing~\cite{HOTI21} have been also evaluated but not reported in this paper for brevity.
	Due to the nature of Polarized routing, the differences between the resource allocation strategies are less pronounced, but the conclusions drawn are consistent with those obtained with Omni-WAR.

	Two different frameworks are considered: interference analysis and single-application scaling.
	In the \textsl{interference analysis scenario}, one target application is allocated in the network and its completion time is measured.
	Background noise is generated with the remaining endpoints, sending traffic as a Random Permutation.
	In the \textsl{single-application scaling} scenario, $k$ replicas of the same application are launched and the makespan (time until all replicas complete) is measured.
	The replica count increases from $k=1$ up to the maximum that fits in the 512-server system.
	This maximum size is 8, 4, and 2 for applications of 64, 128, and 256 processes, respectively.

	\subsection{Analyzed scenarios}

	Traffics with different characteristics can lead to different performance results for the same resource allocation strategy.
	In addition, resource allocation strategies are very sensitive to other network characteristics, such as the number of virtual channels used.
	So, to ensure that the conclusions drawn are robust and not specific to a particular configuration, the performance of the different strategies is evaluated under different scenarios.

	\subsubsection{Static traffic analysis}

	Simulations with static traffic are carried out to analyze the performance of the different strategies under controlled conditions.
	Furthermore, simulations using both MIN routing and Omni-WAR are carried out to confirm that the different strategies yield different performance results, and that the choice of the allocation function is relevant for the performance of applications.
	So, two scenarios are considered: the first one is a single-application scaling using MIN routing and employing the traffic patterns Uniform and Random Permutation.	Secondly, an interference analysis is performed using Omni-WAR routing, with Uniform and Random Switch Permutation traffic patterns. In this case, the background noise is generated using a Random Permutation.
In both scenarios, applications are conformed by 64 processes.
	The use of MIN routing for the interference analysis scenario is not considered, as it would not be able to handle the adversarial traffic patterns nor load balance the load.

	\subsubsection{Kernel traffic analysis}

	The performance of the different resource allocation strategies is evaluated using the communication kernels described.
	Two scenarios are considered, as before. The first one is a single-application scaling of the kernels and the second,
an interference analysis, in which the performance of one application is measured under the presence of a global background noise generated by a Random Permutation.
	For both scenarios the size of the application varies from 64 to 256 processes, and the makespan is measured for up to 8 replicas of the application, depending on its size.
	However, to draw specific conclusions not all the simulations are reported, but only a subset are shown to illustrate the main points of the analysis, and a summary of the results is provided in the form of tables.

	\subsubsection{Fabric partitioning}

	To remove head of line (HoL) blocking and isolate partitions from each other, each partition is given its own set of virtual channels.
	Thus, each partition is provided with 4 unique virtual channels.
	So, for example, in the case of 8 replicas of a 64-process application, which occupies all the 512 endpoints, there are 8 partitions and each partition has its own set of 4 virtual channels, giving a total of 32 virtual channels per switch port.

	\section{Experimental Results}\label{sec:experiments}

	This section is organized into three different subsections.
	Subsection \ref{subsec:minrouting} is devoted to show the relevance of the resource allocation strategies, by comparing their performance under MIN and Omni-WAR routing with simple synthetic traffic patterns.
	Subsection \ref{subsec:kernel-traffic-analysis} is based on the communication kernels and uses Omni-WAR routing.
It has two parts, the first one considering the escalation of a unique application,  and the second, showing the behavior of one application when it experiences interference from a global background noise.
Subsection \ref{sec:vcs} is devoted to show equivalent results but using more virtual channels.

	\subsection{Static traffic analysis}\label{subsec:minrouting}

	\bgroup

\newcommand{\kindseparator}{\hskip 0ex{}}
\pgfplotsset{compat=newest}
\makeatletter
\tikzset{nomorepostaction/.code=\let\tikz@postactions\pgfutil@empty}
\makeatother
\pgfplotsset{minor grid style={dashed,very thin, color=blue!15}}
\pgfplotsset{major grid style={very thin, color=black!30}}
\pgfplotsset{
	automatically generated axis/.style={
		height=105pt,%
		width=200pt,%
		scaled ticks=true,
		xticklabel style={font=\tiny,/pgf/number format/.cd, fixed,/tikz/.cd},%
		yticklabel style={font=\tiny,/pgf/number format/.cd, fixed,/tikz/.cd},%
		x label style={at={(ticklabel cs:0.5, -5pt)},name={x label},anchor=north,font=\scriptsize},
		y label style={at={(ticklabel cs:0.5, -5pt)},name={y label},anchor=south,font=\scriptsize},
		every axis title/.append style={at={(1,0.1)},font=\small,anchor=north east},
	},
	automatically generated symbolic/.style={
		height=105pt,
		width=500pt,
		xticklabel style={font=\tiny,rotate=90},
		yticklabel style={font=\tiny,/pgf/number format/.cd, fixed,/tikz/.cd},%
		x label style={at={(ticklabel cs:0.5, -5pt)},name={x label},anchor=north,font=\scriptsize},
		y label style={at={(ticklabel cs:0.5, -5pt)},name={y label},anchor=south,font=\scriptsize},
	},
	first kind/.style={
		legend style={font=\scriptsize,fill=none},
		legend columns=3,legend cell align=left,
	},
	posterior kind/.style={
		legend style={draw=none},
	},
}
\def\timetickcode{%
	\pgfkeys{/pgf/fpu,/pgf/fpu/output format=fixed}%
	\pgfmathparse{\tick}%
	\edef\tmp{\pgfmathresult}%
	\pgfmathtruncatemacro\seconds{\tmp-60*floor(\tmp/60)}%
	\pgfmathtruncatemacro\tmp{(\tmp - \seconds)/60}%
	\pgfmathtruncatemacro\minutes{\tmp-60*floor(\tmp/60)}%
	\pgfmathtruncatemacro\tmp{(\tmp - \minutes)/60}%
	\pgfmathtruncatemacro\hours{\tmp-24*floor(\tmp/24)}%
	\pgfmathtruncatemacro\days{(\tmp - \hours)/24}%
	\ifnum\days=0%
		{\tiny\hours:\minutes:\seconds}%
	\else%
		{\tiny\days-\hours:\minutes}%
	\fi%
}
\def\memorytickcode{%
	\pgfkeys{/pgf/fpu,/pgf/fpu/output format=fixed}%
	\pgfmathtruncatemacro\unitcase{log2(\tick+0.001)/10+1.1}%
	\pgfmathparse{\tick / pow(1024,\unitcase-1)}%
	\pgfmathprintnumber{\pgfmathresult}%
	\ifcase\unitcase B%
		\or KB%
		\or MB%
		\or GB%
		\or TB%
		\else +1024,\pgfmathprintnumber{unitcase}B%
	\fi%
}
\tikzset{
	automatically generated plot/.style={
		/pgfplots/error bars/x dir=both,
		/pgfplots/error bars/y dir=both,
		/pgfplots/error bars/x explicit,
		/pgfplots/error bars/y explicit,
		/pgfplots/error bars/error bar style={ultra thin,solid},
		/tikz/mark options={solid},
		very thick,
		mark size=2.5pt,
	},
	automatically generated bar plot/.style={
		/pgfplots/error bars/y dir=both,
		/pgfplots/error bars/y explicit,
	},
	automatically generated boxplot/.style={
	},
	x time ticks/.style={
		/pgfplots/scaled x ticks=false,
		/pgfplots/xticklabel={\timetickcode},
	},
	y time ticks/.style={
		/pgfplots/scaled y ticks=false,
		/pgfplots/yticklabel={\timetickcode}
	},
	x memory ticks from kilobytes/.style={
		/pgfplots/scaled x ticks=false,
		/pgfplots/xticklabel={\memorytickcode}
	},
	y memory ticks from kilobytes/.style={
		/pgfplots/scaled y ticks=false,
		/pgfplots/yticklabel={\memorytickcode}
	},
}

\newcommand\captionprologue{X: }
\newcommand\experimenttitle{15-pruebas-randperm-MIN/load_escalation_mappings_2.pdf (all 168 done)}
\newcommand\experimentheader{\tiny 15-pruebas-randperm-MIN:load\_escalation\_mappings\_2.pdf (all 168 done)\\pdflatex on \today\\version=heads/alex-dev-dirty-8e2a39f34b812569e780c03ed6a7e158aa4b09bc(0.6.3)}
\tikzset{xxxxRandomxxxxx/.style={automatically generated plot,red,solid,mark=o}}
\tikzset{xxxxRandomxxxxxbar/.style={automatically generated bar plot,fill=red!20,postaction={pattern=horizontal lines},}}
\tikzset{xxxxRandomxxxxxboxplot/.style={automatically generated boxplot,fill=red!20,every path/.style={postaction={nomorepostaction,pattern=horizontal lines},}}}
\tikzset{xxxxRowsxxxxx/.style={automatically generated plot,green,dashed,mark=square}}
\tikzset{xxxxRowsxxxxxbar/.style={automatically generated bar plot,fill=green!20,postaction={pattern=grid},}}
\tikzset{xxxxRowsxxxxxboxplot/.style={automatically generated boxplot,fill=green!20,every path/.style={postaction={nomorepostaction,pattern=grid},}}}
\tikzset{xxxxRRxswitchesxxxxx/.style={automatically generated plot,blue,dotted,mark=triangle}}
\tikzset{xxxxRRxswitchesxxxxxbar/.style={automatically generated bar plot,fill=blue!20,postaction={pattern=crosshatch},}}
\tikzset{xxxxRRxswitchesxxxxxboxplot/.style={automatically generated boxplot,fill=blue!20,every path/.style={postaction={nomorepostaction,pattern=crosshatch},}}}
\tikzset{xxxxRVIIIxRIVxRIIxxxxx/.style={automatically generated plot,black,dash dot,mark=star}}
\tikzset{xxxxRVIIIxRIVxRIIxxxxxbar/.style={automatically generated bar plot,fill=black!20,postaction={pattern=dots},}}
\tikzset{xxxxRVIIIxRIVxRIIxxxxxboxplot/.style={automatically generated boxplot,fill=black!20,every path/.style={postaction={nomorepostaction,pattern=dots},}}}
\tikzset{xxxxLxtilesxxxxx/.style={automatically generated plot,violet,solid,mark=diamond}}
\tikzset{xxxxLxtilesxxxxxbar/.style={automatically generated bar plot,fill=violet!20,postaction={pattern=north east lines},}}
\tikzset{xxxxLxtilesxxxxxboxplot/.style={automatically generated boxplot,fill=violet!20,every path/.style={postaction={nomorepostaction,pattern=north east lines},}}}
\tikzset{xxxxDiagonalsxxxxx/.style={automatically generated plot,orange,dashed,mark=Mercedes star flipped}}
\tikzset{xxxxDiagonalsxxxxxbar/.style={automatically generated bar plot,fill=orange!20,postaction={pattern=vertical lines},}}
\tikzset{xxxxDiagonalsxxxxxboxplot/.style={automatically generated boxplot,fill=orange!20,every path/.style={postaction={nomorepostaction,pattern=vertical lines},}}}
\tikzset{xxxxRandomxSwitchxSelectionxxxxx/.style={automatically generated plot,lightgray,dotted,mark=o}}
\tikzset{xxxxRandomxSwitchxSelectionxxxxxbar/.style={automatically generated bar plot,fill=lightgray!20,postaction={pattern=horizontal lines},}}
\tikzset{xxxxRandomxSwitchxSelectionxxxxxboxplot/.style={automatically generated boxplot,fill=lightgray!20,every path/.style={postaction={nomorepostaction,pattern=horizontal lines},}}}

\def\xxxxDiagonalsxxxxxtext{Diagonal}
\def\xxxxRandomxSwitchxSelectionxxxxxtext{Random Switch}
\def\xxxxRVIIIxRIVxRIIxxxxxtext{Rectangular}
\def\xxxxLxtilesxxxxxtext{L-shape}
\def\xxxxRandomxxxxxtext{Random Endpoint}

\begin{experimentfiguressmall}%
	\centering%
	\tikzpicturedependsonfile{externalized-plots/external-RIRVxpruebasxrandpermxMIN-throughput-selectorxxxxTMcxNaturalxxxxxRandomPermutationxxxxxMINxxxxx-kind0.md5}%
	\tikzsetnextfilename{externalized-legends/legend-RIRVxpruebasxrandpermxMIN-throughput-xxxxTMcxNaturalxxxxxRandomPermutationxxxxxMINxxxxx}%
	\pgfplotslegendfromname{legend-RIRVxpruebasxrandpermxMIN-throughput-xxxxTMcxNaturalxxxxxRandomPermutationxxxxxMINxxxxx}\\
	\tikzsetnextfilename{externalized-plots/external-RIRVxpruebasxrandpermxMIN-throughput-selectorxxxxTMcxNaturalxxxxxRandomPermutationxxxxxMINxxxxx-kind0}%
	\begin{tikzpicture}[baseline,remember picture]
	\begin{axis}[
		automatically generated axis,
		first kind,,
		legend to name=legend-RIRVxpruebasxrandpermxMIN-throughput-xxxxTMcxNaturalxxxxxRandomPermutationxxxxxMINxxxxx,
		title={Random permutation},
		ymin=0,xmin=0,%
		ymajorgrids=true,
		yminorgrids=true,
		xmajorgrids=true,
		mark options=solid,
		minor y tick num=4,
		xlabel={Allocated Kernels},
		ylabel={Cycles},
	]
\addplot[xxxxDiagonalsxxxxx] coordinates{(1,133846.67)  (2,142668)  (3,160321.33) (4,216737.33)  (5,220490.67)  (6,224478.67)  (7,242240)  (8,251030.67) };\addlegendentry{\xxxxDiagonalsxxxxxtext}
\addplot[xxxxRandomxSwitchxSelectionxxxxx] coordinates{(1,209022.67)  (2,232744) (3,239624) (4,273858.66)  (5,338173.34)  (6,306024)  (7,338212)  (8,326461.34) };\addlegendentry{\xxxxRandomxSwitchxSelectionxxxxxtext}
\addplot[xxxxRVIIIxRIVxRIIxxxxx] coordinates{(1,424461.34)  (2,426040)  (3,442738.66)  (4,475780)  (5,494504) (6,494504) (7,495464) (8,494205.34)};\addlegendentry{\xxxxRVIIIxRIVxRIIxxxxxtext}
\addplot[xxxxLxtilesxxxxx] coordinates{(1,267080)  (2,267858.66)  (3,290162.66)  (4,377533.34)  (5,378098.66)  (6,378546.66)  (7,378386.66)  (8,378130.66) };\addlegendentry{\xxxxLxtilesxxxxxtext}
\addplot[xxxxRowsxxxxx] coordinates{(1,266248)  (2,266248)  (3,266248)  (4,346120)  (5,346120)  (6,346120)  (7,346120)  (8,346120) };\addlegendentry{\xxxxRowsxxxxxtext}
\addplot[xxxxRandomxxxxx] coordinates{(1,146440)  (2,160788)  (3,178706.67)  (4,170962.67) (5,193021.33) (6,213117.33)  (7,239273.33)  (8,253159.33) };\addlegendentry{\xxxxRandomxxxxxtext}
\addplot[xxxxRRxswitchesxxxxx] coordinates{(1,79912) (2,140520)  (3,160306.67) (4,182549.33) (5,195906.67) (6,211086.67)  (7,223012) (8,246777.33)};\addlegendentry{\xxxxRRxswitchesxxxxxtext}
	\end{axis}
	\end{tikzpicture}%

	\tikzsetnextfilename{externalized-plots/external-RIRIIIbxpruebasxtrafficxuniformexMIN-throughput-selectorxxxxTMcxNaturalxxxxxUniformxxxxxMINxxxxx-kind0}%
	\begin{tikzpicture}[baseline,remember picture]
		\begin{axis}[
			automatically generated axis,
			first kind,,
			legend to name=legend-RIRIIIbxpruebasxtrafficxuniformexMIN-throughput-xxxxTMcxNaturalxxxxxUniformxxxxxMINxxxxx,
			title={Uniform},
			ymin=0,xmin=0,%
			ymajorgrids=true,
			yminorgrids=true,
			xmajorgrids=true,
			mark options=solid,
			minor y tick num=4,
			xlabel={Allocated Kernels},
			ylabel={Cycles},
		]
			\addplot[xxxxDiagonalsxxxxx] coordinates{(1,90161.336) (2,90334) (3,93733.336) (4,93807.336) (5,93396) (6,93718.664) (7,92788.664) (8,92356.664)};\addlegendentry{\xxxxDiagonalsxxxxxtext}
			\addplot[xxxxRandomxSwitchxSelectionxxxxx] coordinates{(1,125279.336) (2,154864)  (3,176984)  (4,179784)  (5,178131.33)  (6,179108)  (7,189400.67)  (8,202664.67) };\addlegendentry{\xxxxRandomxSwitchxSelectionxxxxxtext}
			\addplot[xxxxRVIIIxRIVxRIIxxxxx] coordinates{(1,325699.34) (2,327668.66) (3,327796.66) (4,328471.34) (5,328754.66) (6,328742.66) (7,329746.66) (8,329461.34)};\addlegendentry{\xxxxRVIIIxRIVxRIIxxxxxtext}
			\addplot[xxxxLxtilesxxxxx] coordinates{(1,103862.664) (2,102259.336) (3,103542) (4,103288.664) (5,104886.664) (6,103214.664) (7,106050) (8,103937.336)};\addlegendentry{\xxxxLxtilesxxxxxtext}
			\addplot[xxxxRowsxxxxx] coordinates{(1,91986) (2,94519.336) (3,94616) (4,94210) (5,93756) (6,94743.336) (7,94730.664) (8,94249.336)};\addlegendentry{\xxxxRowsxxxxxtext}
			\addplot[xxxxRandomxxxxx] coordinates{(1,101128.664)  (2,102042.664)  (3,105534)  (4,104896.664) (5,103703.336) (6,100990.664) (7,100728) (8,98634.664)};\addlegendentry{\xxxxRandomxxxxxtext}
			\addplot[xxxxRRxswitchesxxxxx] coordinates{(1,84246.664) (2,86649.336) (3,87493.336) (4,90476) (5,91618) (6,92441.336) (7,91141.336) (8,92880)};\addlegendentry{\xxxxRRxswitchesxxxxxtext}
		\end{axis}
	\end{tikzpicture}%
	\caption{Makespan for load escalation of 64-processes kernels with MIN routing, allocating in the network from 1 to 8 replicas of the same app in each chart.}%
	\label{fig:minrouting}%
\end{experimentfiguressmall}

\egroup

	\bgroup
\pgfplotsset{compat=newest}
\makeatletter
\tikzset{nomorepostaction/.code=\let\tikz@postactions\pgfutil@empty}
\makeatother
\pgfplotsset{minor grid style={dashed,very thin, color=blue!15}}
\pgfplotsset{major grid style={very thin, color=black!30}}
\pgfplotsset{
    automatically generated axis/.style={
       height=105pt, width=174pt, scaled ticks=false,
       xticklabel style={font=\tiny,/pgf/number format/.cd, fixed,/tikz/.cd},
       yticklabel style={font=\tiny,/pgf/number format/.cd, fixed,/tikz/.cd},
       x label style={at={(ticklabel cs:0.5, -5pt)},name={x label},anchor=north,font=\scriptsize},
       y label style={at={(ticklabel cs:0.5, -5pt)},name={y label},anchor=south,font=\scriptsize},
    },
    automatically generated symbolic/.style={
       height=125pt, width=245pt,%
       xticklabel style={font=\small},
       yticklabel style={font=\small,/pgf/number format/.cd, fixed,/tikz/.cd},
       x label style={at={(ticklabel cs:0.5, -5pt)},name={x label},anchor=north,font=\footnotesize},
       y label style={at={(ticklabel cs:0.5, -5pt)},name={y label},anchor=south,font=\footnotesize},
    },
    first kind/.style={
		bar legend,
       legend columns=3,
    }
}
\tikzset{
    automatically generated plot/.style={
       /pgfplots/error bars/x dir=both, /pgfplots/error bars/y dir=both,
       /pgfplots/error bars/x explicit, /pgfplots/error bars/y explicit,
       /pgfplots/error bars/error bar style={ultra thin,solid},
       /tikz/mark options={solid},
    },
    automatically generated bar plot/.style={
       /pgfplots/error bars/y dir=both, /pgfplots/error bars/y explicit,
       /pgfplots/error bars/error bar style={line width=7pt,solid,black!40},
       /pgfplots/error bars/error mark=none,
    }
}

\def\xInterxwithxrandomxswitchxpermxtxt{"Inter with random switch perm"}
\def\xInterxwithxuniformxtxt{"Inter with uniform"}
\def\xInterferencedxwithxRandomPermutationxtxt{"Interferenced with RandomPermutation"}
\def\xInterferencedxwithxRandomPermutationx{0}
\def\xNoxInterferredxtxt{"No Interferred"}
\def\uniformtraffic{0}
\def\randomswitchpermtraffic{1}

\def\xxxxDiagonalsxxxxxtext{Diagonal }
\def\xxxxRandomxSwitchxSelectionxxxxxtext{Random Switch }
\def\xxxxRVIIIxRIVxRIIxxxxxtext{Rectangular}
\def\xxxxLxtilesxxxxxtext{L-shape}
\def\xxxxRandomxxxxxtext{Random Endpoint }

\begin{experimentfiguressmall}
    \centering
    \pgfplotslegendfromname{leyenda-random-inter-no-VC}\\
    \begin{tikzpicture}[baseline,remember picture]
    \begin{axis}[
       automatically generated symbolic,
       xtick={0, 1},
       xticklabels = {{Uniform}, {Random Switch Perm}},
       ybar,bar width=8pt,enlarge x limits=0.6,
       first kind,,
       legend to name=leyenda-random-inter-no-VC,
       ymin=0,
       ymajorgrids=true, yminorgrids=true, xmajorgrids=true,
       mark options=solid, minor y tick num=4,
       ylabel={Cycles},
    ]
        \addplot[xxxxDiagonalsxxxxxbar] coordinates{  (\uniformtraffic,92988.80) -=(0, 0) +=(0, 13926.40) (\randomswitchpermtraffic,80116.40) -=(0, 0) +=(0, 84777.20)};\addlegendentry{\xxxxDiagonalsxxxxxtext}
        \addplot[xxxxRandomxSwitchxSelectionxxxxxbar] coordinates{  (\uniformtraffic,94876.80) -=(0, 0) +=(0, 32616.00) (\randomswitchpermtraffic,92193.20) -=(0, 0) +=(0, 173746.00)};\addlegendentry{\xxxxRandomxSwitchxSelectionxxxxxtext}
        \addplot[xxxxRVIIIxRIVxRIIxxxxxbar] coordinates{  (\uniformtraffic,105497.60) -=(0, 0) +=(0, 35115.20) (\randomswitchpermtraffic,110809.60) -=(0, 0) +=(0, 214468.80)};\addlegendentry{\xxxxRVIIIxRIVxRIIxxxxxtext}
        \addplot[xxxxLxtilesxxxxxbar] coordinates{  (\uniformtraffic,111478.00) -=(0, 0) +=(0, 36607.60) (\randomswitchpermtraffic,117831.20) -=(0, 0) +=(0, 198742.00)};\addlegendentry{\xxxxLxtilesxxxxxtext}
        \addplot[xxxxRowsxxxxxbar] coordinates{  (\uniformtraffic,112452.00) -=(0, 0) +=(0, 70423.60)  (\randomswitchpermtraffic,148258.40) -=(0, 0) +=(0, 136556.40)};\addlegendentry{\xxxxRowsxxxxxtext}
        \addplot[xxxxRandomxxxxxbar] coordinates{  (\uniformtraffic,92352.00) -=(0, 0) +=(0, 44069.60) (\randomswitchpermtraffic,79962.40) -=(0, 0) +=(0, 275143.60) };\addlegendentry{\xxxxRandomxxxxxtext}
        \addplot[xxxxRRxswitchesxxxxxbar] coordinates{  (\uniformtraffic,93045.60) -=(0, 0) +=(0, 55493.60) (\randomswitchpermtraffic,79880.00) -=(0, 0) +=(0, 305193.20) };\addlegendentry{\xxxxRRxswitchesxxxxxtext}
    \end{axis}
    \end{tikzpicture}
    \caption{Makespan of 64-process kernels with Omni-WAR routing. The light-colored bars show the time running isolated and the dark part shows the extra time taken due to the interference.}
    \label{fig:omni-basic-traffic-interference-no-vc}
\end{experimentfiguressmall}

\egroup

	\bgroup

\pgfplotsset{compat=newest}
\makeatletter
\tikzset{nomorepostaction/.code=\let\tikz@postactions\pgfutil@empty}
\makeatother
\pgfplotsset{minor grid style={dashed,very thin, color=blue!15}}
\pgfplotsset{major grid style={very thin, color=black!30}}
\pgfplotsset{
	automatically generated axis/.style={
		height=120pt,%
		width=185pt,%
		scaled ticks=true,
		xticklabel style={font=\footnotesize,/pgf/number format/.cd, fixed,/tikz/.cd},%
		yticklabel style={font=\footnotesize,/pgf/number format/.cd, fixed,/tikz/.cd},%
		x label style={at={(ticklabel cs:0.5, -5pt)},name={x label},anchor=north,font=\small},
		y label style={at={(ticklabel cs:0.5, -5pt)},name={y label},anchor=south,font=\small},
		every axis title/.append style={at={(1,0.1)},font=\small,anchor=north east},
	},
	automatically generated symbolic/.style={
		height=105pt,
		width=500pt,
		xticklabel style={font=\tiny,rotate=90},
		yticklabel style={font=\tiny,/pgf/number format/.cd, fixed,/tikz/.cd},%
		x label style={at={(ticklabel cs:0.5, -5pt)},name={x label},anchor=north,font=\scriptsize},
		y label style={at={(ticklabel cs:0.5, -5pt)},name={y label},anchor=south,font=\scriptsize},
	},
	first kind/.style={
		legend style={font=\small,fill=none},
		legend columns=1,legend cell align=left,
	},
	posterior kind/.style={
		legend style={draw=none},
	},
}
\def\timetickcode{%
	\pgfkeys{/pgf/fpu,/pgf/fpu/output format=fixed}%
	\pgfmathparse{\tick}%
	\edef\tmp{\pgfmathresult}%
	\pgfmathtruncatemacro\seconds{\tmp-60*floor(\tmp/60)}%
	\pgfmathtruncatemacro\tmp{(\tmp - \seconds)/60}%
	\pgfmathtruncatemacro\minutes{\tmp-60*floor(\tmp/60)}%
	\pgfmathtruncatemacro\tmp{(\tmp - \minutes)/60}%
	\pgfmathtruncatemacro\hours{\tmp-24*floor(\tmp/24)}%
	\pgfmathtruncatemacro\days{(\tmp - \hours)/24}%
	\ifnum\days=0%
		{\tiny\hours:\minutes:\seconds}%
	\else%
		{\tiny\days-\hours:\minutes}%
	\fi%
}
\def\memorytickcode{%
	\pgfkeys{/pgf/fpu,/pgf/fpu/output format=fixed}%
	\pgfmathtruncatemacro\unitcase{log2(\tick+0.001)/10+1.1}%
	\pgfmathparse{\tick / pow(1024,\unitcase-1)}%
	\pgfmathprintnumber{\pgfmathresult}%
	\ifcase\unitcase B%
		\or KB%
		\or MB%
		\or GB%
		\or TB%
		\else +1024,\pgfmathprintnumber{unitcase}B%
	\fi%
}
\tikzset{
	automatically generated plot/.style={
		/pgfplots/error bars/x dir=both,
		/pgfplots/error bars/y dir=both,
		/pgfplots/error bars/x explicit,
		/pgfplots/error bars/y explicit,
		/pgfplots/error bars/error bar style={ultra thin,solid},
		/tikz/mark options={solid},
		very thick,
		mark size=2.5pt,
	},
	automatically generated bar plot/.style={
		/pgfplots/error bars/y dir=both,
		/pgfplots/error bars/y explicit,
	},
	automatically generated boxplot/.style={
	},
	x time ticks/.style={
		/pgfplots/scaled x ticks=false,
		/pgfplots/xticklabel={\timetickcode},
	},
	y time ticks/.style={
		/pgfplots/scaled y ticks=false,
		/pgfplots/yticklabel={\timetickcode}
	},
	x memory ticks from kilobytes/.style={
		/pgfplots/scaled x ticks=false,
		/pgfplots/xticklabel={\memorytickcode}
	},
	y memory ticks from kilobytes/.style={
		/pgfplots/scaled y ticks=false,
		/pgfplots/yticklabel={\memorytickcode}
	},
}

\newcommand\captionprologue{X: }
\newcommand\experimenttitle{05b-pruebas-expr-todo/load_escalation_mappings_2.pdf (all 3360 done)}
\newcommand\experimentheader{\tiny 05b-pruebas-expr-todo:load\_escalation\_mappings\_2.pdf (all 3360 done)\\pdflatex on \today\\version=heads/alex-stable-dirty-013e5c1afcea726af926629838ffa011555fb4ed(0.6.3)}

\tikzset{xxxxRandomxxxxx/.style={automatically generated plot,red,solid,mark=o}}
\tikzset{xxxxRandomxxxxxbar/.style={automatically generated bar plot,fill=red!20,postaction={pattern=horizontal lines},}}
\tikzset{xxxxRandomxxxxxboxplot/.style={automatically generated boxplot,fill=red!20,every path/.style={postaction={nomorepostaction,pattern=horizontal lines},}}}
\tikzset{xxxxRowsxxxxx/.style={automatically generated plot,green,dashed,mark=square}}
\tikzset{xxxxRowsxxxxxbar/.style={automatically generated bar plot,fill=green!20,postaction={pattern=grid},}}
\tikzset{xxxxRowsxxxxxboxplot/.style={automatically generated boxplot,fill=green!20,every path/.style={postaction={nomorepostaction,pattern=grid},}}}
\tikzset{xxxxRRxswitchesxxxxx/.style={automatically generated plot,blue,dotted,mark=triangle}}
\tikzset{xxxxRRxswitchesxxxxxbar/.style={automatically generated bar plot,fill=blue!20,postaction={pattern=crosshatch},}}
\tikzset{xxxxRRxswitchesxxxxxboxplot/.style={automatically generated boxplot,fill=blue!20,every path/.style={postaction={nomorepostaction,pattern=crosshatch},}}}
\tikzset{xxxxRVIIIxRIVxRIIxxxxx/.style={automatically generated plot,black,dash dot,mark=star}}
\tikzset{xxxxRVIIIxRIVxRIIxxxxxbar/.style={automatically generated bar plot,fill=black!20,postaction={pattern=dots},}}
\tikzset{xxxxRVIIIxRIVxRIIxxxxxboxplot/.style={automatically generated boxplot,fill=black!20,every path/.style={postaction={nomorepostaction,pattern=dots},}}}
\tikzset{xxxxLxtilesxxxxx/.style={automatically generated plot,violet,solid,mark=diamond}}
\tikzset{xxxxLxtilesxxxxxbar/.style={automatically generated bar plot,fill=violet!20,postaction={pattern=north east lines},}}
\tikzset{xxxxLxtilesxxxxxboxplot/.style={automatically generated boxplot,fill=violet!20,every path/.style={postaction={nomorepostaction,pattern=north east lines},}}}
\tikzset{xxxxDiagonalsxxxxx/.style={automatically generated plot,orange,dashed,mark=Mercedes star flipped}}
\tikzset{xxxxDiagonalsxxxxxbar/.style={automatically generated bar plot,fill=orange!20,postaction={pattern=vertical lines},}}
\tikzset{xxxxDiagonalsxxxxxboxplot/.style={automatically generated boxplot,fill=orange!20,every path/.style={postaction={nomorepostaction,pattern=vertical lines},}}}
\tikzset{xxxxRandomxSwitchxSelectionxxxxx/.style={automatically generated plot,lightgray,dotted,mark=o}}
\tikzset{xxxxRandomxSwitchxSelectionxxxxxbar/.style={automatically generated bar plot,fill=lightgray!20,postaction={pattern=horizontal lines},}}
\tikzset{xxxxRandomxSwitchxSelectionxxxxxboxplot/.style={automatically generated boxplot,fill=lightgray!20,every path/.style={postaction={nomorepostaction,pattern=horizontal lines},}}}

\begin{experimentfigure}%
	\centering%
	\noindent\vbox{\halign{\hfil#&\hfil#&\hfil#\cr
	\tikzsetnextfilename{externalized-plots/external-RRVbxpruebasxexprxtodo-throughput-selectorxxxxTMcxRandomxxxxxAllxreducexxxxxOmnidimensionalxxxxx-kind0}%
	\begin{tikzpicture}[baseline,remember picture]
	\begin{axis}[
		automatically generated axis,
		first kind,,
		legend to name=legend-RRVbxpruebasxexprxtodo-throughput-xxxxTMcxRandomxxxxxAllxreducexxxxxOmnidimensionalxxxxx,
		title={All-reduce},
		ymin=30000,xmin=0,%
		ymajorgrids=true,
		yminorgrids=true,
		xmajorgrids=true,
		mark options=solid,
		minor y tick num=4,
		xlabel={Allocated Kernels},
		ylabel={Cycles},
	]
		\addplot[xxxxDiagonalsxxxxx] coordinates{(1,92656.664) (2,97209.336) (3,98409.336) (4,99340.664) (5,101764.664) (6,100971.336) (7,134982.67)  (8,140490.67)};\addlegendentry{\xxxxDiagonalsxxxxxtext}
		\addplot[xxxxRandomxSwitchxSelectionxxxxx] coordinates{(1,92629.336) (2,93552.664) (3,97906) (4,96060.664) (5,106092)  (6,119232)  (7,130252) (8,138125.33)};\addlegendentry{\xxxxRandomxSwitchxSelectionxxxxxtext}
		\addplot[xxxxRVIIIxRIVxRIIxxxxx] coordinates{(1,90650) (2,100461.336) (3,105324.664) (4,116109.336)  (5,132706)  (6,140708.67) (7,139390.67) (8,145326)};\addlegendentry{\xxxxRVIIIxRIVxRIIxxxxxtext}
		\addplot[xxxxLxtilesxxxxx] coordinates{(1,114044)  (2,121111.336) (3,121339.336)  (4,129232) (5,132481.33) (6,131805.33) (7,152276)  (8,151256) };\addlegendentry{\xxxxLxtilesxxxxxtext}
		\addplot[xxxxRowsxxxxx] coordinates{(1,147920) (2,147547.33) (3,152757.33) (4,151848) (5,151168.67) (6,150620.67) (7,153392) (8,152683.33)};\addlegendentry{\xxxxRowsxxxxxtext}
		\addplot[xxxxRandomxxxxx] coordinates{(1,82974) (2,92070.664) (3,94187.336) (4,96484.664) (5,97431.336) (6,115012)  (7,128604.664) (8,136301.33)};\addlegendentry{\xxxxRandomxxxxxtext}
		\addplot[xxxxRRxswitchesxxxxx] coordinates{(1,73016)  (2,93122.664) (3,95164.664) (4,95533.336) (5,96393.336) (6,108564) (7,126345.336) (8,133552.67)};\addlegendentry{\xxxxRRxswitchesxxxxxtext}
	\end{axis}
	\end{tikzpicture}&%
	\tikzsetnextfilename{externalized-plots/external-RRVbxpruebasxexprxtodo-throughput-selectorxxxxTMcxRandomxxxxxAllRIIAllxxxxxOmnidimensionalxxxxx-kind0}%
	\begin{tikzpicture}[baseline,remember picture]
	\begin{axis}[
		automatically generated axis,
		first kind,,
		legend to name=legend-RRVbxpruebasxexprxtodo-throughput-xxxxTMcxRandomxxxxxAllRIIAllxxxxxOmnidimensionalxxxxx,
		title={All2All},
		ymin=30000,xmin=0,%
		ymajorgrids=true,
		yminorgrids=true,
		xmajorgrids=true,
		mark options=solid,
		minor y tick num=4,
		xlabel={Allocated Kernels},
		ylabel={Cycles},
	]
\addplot[xxxxDiagonalsxxxxx] coordinates{(1,129101.336) (2,129132) (3,129163.336) (4,129190.664) (5,129202) (6,129224.664) (7,129268.664) (8,134188.67)};\addlegendentry{\xxxxDiagonalsxxxxxtext}
\addplot[xxxxRandomxSwitchxSelectionxxxxx] coordinates{(1,129194.664) (2,129222) (3,129245.336) (4,129251.336) (5,129268) (6,129307.336) (7,167965.33) (8,176211.33) };\addlegendentry{\xxxxRandomxSwitchxSelectionxxxxxtext}
\addplot[xxxxRVIIIxRIVxRIIxxxxx] coordinates{(1,129224) (2,129226.664) (3,129233.336) (4,129256) (5,153898) (6,161031.33) (7,167248) (8,173214.67)};\addlegendentry{\xxxxRVIIIxRIVxRIIxxxxxtext}
\addplot[xxxxLxtilesxxxxx] coordinates{(1,129193.336) (2,129194) (3,129208.664) (4,151887.33) (5,156469.33) (6,158873.33) (7,165360) (8,176031.33)};\addlegendentry{\xxxxLxtilesxxxxxtext}
\addplot[xxxxRowsxxxxx] coordinates{(1,135465.33) (2,135404.67) (3,135570.67) (4,136176) (5,135933.33) (6,136927.33) (7,136459.33) (8,136622)};\addlegendentry{\xxxxRowsxxxxxtext}
\addplot[xxxxRandomxxxxx] coordinates{(1,129110.664) (2,129132) (3,129144) (4,129165.336) (5,129182.664) (6,129220) (7,129264) (8,150379.33)};\addlegendentry{\xxxxRandomxxxxxtext}
\addplot[xxxxRRxswitchesxxxxx] coordinates{(1,129068) (2,129101.336) (3,129128) (4,129143.336) (5,129162.664) (6,129193.336) (7,129246.664) (8,146590)};\addlegendentry{\xxxxRRxswitchesxxxxxtext}
	\end{axis}
	\end{tikzpicture}&%
		\tikzsetnextfilename{externalized-plots/external-RRVbxpruebasxexprxtodo-throughput-selectorxxxxTMcxRandomxxxxxRandomInvolutionxxxxxOmnidimensionalxxxxx-kind0}%
	\begin{tikzpicture}[baseline,remember picture]
	\begin{axis}[
		automatically generated axis,
		first kind,,
		legend to name=legend-RRVbxpruebasxexprxtodo-throughput-xxxxTMcxRandomxxxxxRandomInvolutionxxxxxOmnidimensionalxxxxx,
		title={RandomInvolution},
		ymin=30000,xmin=0,%
		ymajorgrids=true,
		yminorgrids=true,
		xmajorgrids=true,
		mark options=solid,
		minor y tick num=4,
		xlabel={Allocated Kernels},
		ylabel={Cycles},
	]
\addplot[xxxxDiagonalsxxxxx] coordinates{(1,80172) (2,80192) (3,80236.664) (4,80202.664) (5,80236) (6,105849.336) (7,140222.67) (8,150886.67)};\addlegendentry{\xxxxDiagonalsxxxxxtext}
\addplot[xxxxRandomxSwitchxSelectionxxxxx] coordinates{(1,80139.336) (2,80141.336) (3,80184) (4,82443.336) (5,111896)  (6,130351.336)  (7,148694.67) (8,163940)};\addlegendentry{\xxxxRandomxSwitchxSelectionxxxxxtext}
\addplot[xxxxRVIIIxRIVxRIIxxxxx] coordinates{(1,80064) (2,101056)  (3,105741.336)  (4,119033.336)  (5,146836.67)  (6,165860)  (7,169888.67)  (8,173424.67)};\addlegendentry{\xxxxRVIIIxRIVxRIIxxxxxtext}
\addplot[xxxxLxtilesxxxxx] coordinates{(1,87070)  (2,126055.336) (3,133572)  (4,139532.67)  (5,146206.67) (6,149757.33)  (7,157750.67)  (8,164676)};\addlegendentry{\xxxxLxtilesxxxxxtext}
\addplot[xxxxRowsxxxxx] coordinates{(1,173451.33)  (2,181551.33) (3,178652.67) (4,183936) (5,183829.33) (6,176213.33) (7,183976) (8,181966)};\addlegendentry{\xxxxRowsxxxxxtext}
\addplot[xxxxRandomxxxxx] coordinates{(1,79966.664) (2,80026) (3,80074) (4,80105.336) (5,80137.336) (6,124919.336) (7,148362) (8,158832)};\addlegendentry{\xxxxRandomxxxxxtext}
\addplot[xxxxRRxswitchesxxxxx] coordinates{(1,79912) (2,79990.664) (3,80066.664) (4,80094.664) (5,80106.664) (6,133398) (7,149783.33) (8,164120.67)};\addlegendentry{\xxxxRRxswitchesxxxxxtext}
	\end{axis}
	\end{tikzpicture}\cr%
	\tikzsetnextfilename{externalized-plots/external-RRVbxpruebasxexprxtodo-throughput-selectorxxxxTMcxRandomxxxxxStencilxmanhattanxxxxxOmnidimensionalxxxxx-kind0}%
	\begin{tikzpicture}[baseline,remember picture]
	\begin{axis}[
		automatically generated axis,
		first kind,,
		legend to name=legend-RRVbxpruebasxexprxtodo-throughput-xxxxTMcxRandomxxxxxStencilxmanhattanxxxxxOmnidimensionalxxxxx,
		title={Stencil von Neumann},
		ymin=30000,xmin=0,%
		ymajorgrids=true,
		yminorgrids=true,
		xmajorgrids=true,
		mark options=solid,
		minor y tick num=4,
		xlabel={Allocated Kernels},
		ylabel={Cycles},
	]
\addplot[xxxxDiagonalsxxxxx] coordinates{(1,36133.332) (2,37351.332) (3,36754) (4,37596.668) (5,40008) (6,41973.332) (7,44922.668) (8,49899.332)};\addlegendentry{\xxxxDiagonalsxxxxxtext}
\addplot[xxxxRandomxSwitchxSelectionxxxxx] coordinates{(1,36582.668) (2,36817.332) (3,36679.332) (4,38362)  (5,42440.668) (6,43756.668) (7,48950.668) (8,50639.332)};\addlegendentry{\xxxxRandomxSwitchxSelectionxxxxxtext}
\addplot[xxxxRVIIIxRIVxRIIxxxxx] coordinates{(1,36011.332) (2,39606) (3,40986) (4,40845.332) (5,51693.332)  (6,53233.332) (7,55104)  (8,58444.668)};\addlegendentry{\xxxxRVIIIxRIVxRIIxxxxxtext}
\addplot[xxxxLxtilesxxxxx] coordinates{(1,38916.668) (2,44231.332)  (3,51118)  (4,50486.668)  (5,51530.668) (6,52318.668) (7,55775.332) (8,57232.668)};\addlegendentry{\xxxxLxtilesxxxxxtext}
\addplot[xxxxRowsxxxxx] coordinates{(1,48453.332) (2,51457.332) (3,51290) (4,51948.668) (5,51730) (6,52363.332) (7,52763.332) (8,53418)};\addlegendentry{\xxxxRowsxxxxxtext}
\addplot[xxxxRandomxxxxx] coordinates{(1,35916.668) (2,36552) (3,36822.668) (4,36414.668) (5,38154) (6,41004.668) (7,42846.668) (8,46857.332)};\addlegendentry{\xxxxRandomxxxxxtext}
\addplot[xxxxRRxswitchesxxxxx] coordinates{(1,32981.332) (2,34941.332) (3,35651.332) (4,35653.332) (5,36614) (6,40578) (7,42732) (8,46334)};\addlegendentry{\xxxxRRxswitchesxxxxxtext}
	\end{axis}
	\end{tikzpicture}&%
		\tikzsetnextfilename{externalized-plots/external-RRVbxpruebasxexprxtodo-throughput-selectorxxxxTMcxRandomxxxxxStencilxkingxxxxxOmnidimensionalxxxxx-kind0}%
	\begin{tikzpicture}[baseline,remember picture]
	\begin{axis}[
		automatically generated axis,
		first kind,,
		legend to name=legend-RRVbxpruebasxexprxtodo-throughput-xxxxTMcxRandomxxxxxStencilxkingxxxxxOmnidimensionalxxxxx,
		title={Stencil Moore},
		ymin=30000,xmin=0,%
		ymajorgrids=true,
		yminorgrids=true,
		xmajorgrids=true,
		mark options=solid,
		minor y tick num=4,
		xlabel={Allocated Kernels},
		ylabel={Cycles},
	]
\addplot[xxxxDiagonalsxxxxx] coordinates{(1,77580.664) (2,80562.664) (3,84956.664) (4,95916.664) (5,96164.664) (6,98191.336) (7,97825.336) (8,103426)};\addlegendentry{\xxxxDiagonalsxxxxxtext}
\addplot[xxxxRandomxSwitchxSelectionxxxxx] coordinates{(1,78622.664) (2,90271.336) (3,88415.336) (4,98658.664) (5,101598.664) (6,101182.664) (7,109424.664) (8,107698)};\addlegendentry{\xxxxRandomxSwitchxSelectionxxxxxtext}
\addplot[xxxxRVIIIxRIVxRIIxxxxx] coordinates{(1,97260) (2,94946) (3,96530.664) (4,98496) (5,101687.336) (6,105812.664) (7,108424.664) (8,112534)};\addlegendentry{\xxxxRVIIIxRIVxRIIxxxxxtext}
\addplot[xxxxLxtilesxxxxx] coordinates{(1,96634.664)  (2,102942) (3,105068.664) (4,102622) (5,103330) (6,107738.664) (7,107100.664)  (8,115421.336)};\addlegendentry{\xxxxLxtilesxxxxxtext}
\addplot[xxxxRowsxxxxx] coordinates{(1,95172) (2,96522) (3,101192) (4,101822) (5,101700.664) (6,102802) (7,104486.664) (8,102622)};\addlegendentry{\xxxxRowsxxxxxtext}
\addplot[xxxxRandomxxxxx] coordinates{(1,76768) (2,80526.664) (3,89590) (4,91236.664) (5,93348) (6,94889.336) (7,96580.664) (8,97387.336)};\addlegendentry{\xxxxRandomxxxxxtext}
\addplot[xxxxRRxswitchesxxxxx] coordinates{(1,70530.664) (2,77032.664) (3,80399.336) (4,83744) (5,87452.664) (6,91010.664) (7,94901.336) (8,97992)};\addlegendentry{\xxxxRRxswitchesxxxxxtext}
	\end{axis}
	\end{tikzpicture}&%
	\tikzpicturedependsonfile{externalized-plots/external-RRVbxpruebasxexprxtodo-throughput-selectorxxxxTMcxRandomxxxxxAllRIIAllxxxxxOmnidimensionalxxxxx-kind0.md5}%
	\tikzsetnextfilename{externalized-legends/legend-RRVbxpruebasxexprxtodo-throughput-xxxxTMcxRandomxxxxxAllRIIAllxxxxxOmnidimensionalxxxxx}%
	\pgfplotslegendfromname{legend-RRVbxpruebasxexprxtodo-throughput-xxxxTMcxRandomxxxxxAllRIIAllxxxxxOmnidimensionalxxxxx}\cr%
	}}%
	\caption{Makespan for load escalation of 64-processes kernels with Omni-WAR routing, allocating in the network from 1 to 8 replicas of the same app in each chart.}%
	\label{fig:omni_64_random}%
\end{experimentfigure}

\egroup

	In Figure~\ref{fig:minrouting} MIN is used, and it can be observed that the choice of the resource allocation strategy has a significant impact on application performance.
	Rectangles is the worst strategy in both traffic patterns, next Random Switch Selection, L-shapes and Rows follow, and lastly Diagonal, Random Endpoint and Full Spread are the best strategies, with similar performance.
	It can be observed that, due to the asymmetry of the partition, the performance of Random Switch Selection is worse than expected under Uniform traffic pattern.
	Here, the routing algorithm is not able to properly load balance the traffic, and some switches have to handle more traffic than others, which leads to unfairness between the endpoints of the partition and a decrease in performance.
	An ideal routing algorithm with global knowledge of the traffic would be able to balance the traffic across the network and achieve better performance, but MIN is not able to do so.
	The performance of the different strategies is consistent with the Partition Bandwidth values shown in Table~\ref{tbl:analysis}.

	In Figure~\ref{fig:omni-basic-traffic-interference-no-vc} it can be observed the performance of the different strategies for Uniform and Random Switch Permutation with the presence of background noise generated with a Random Permutation. Omni-WAR routing is used.
	As it can be seen, Random and Full Spread are the worst strategies as the ones most troubled by interference and HoL blocking.
	Then Rectangles, L-shapes, Rows and Random Switch Selection follow, with similar performance.
	Finally, Diagonal is the best strategy, as it is more robust to interference than Full-Spread and Random, and maintains a high Partition Bandwidth.

	\subsection{Kernel traffic analysis}\label{subsec:kernel-traffic-analysis}

	\subsubsection{Application escalation}\label{subsec:scalation}

	\bgroup
\pgfplotsset{compat=newest}
\makeatletter
\tikzset{nomorepostaction/.code=\let\tikz@postactions\pgfutil@empty}
\makeatother
\pgfplotsset{minor grid style={dashed,very thin, color=blue!15}}
\pgfplotsset{major grid style={very thin, color=black!30}}
\pgfplotsset{
    automatically generated axis/.style={
        height=105pt, width=174pt, scaled ticks=false,
        xticklabel style={font=\tiny,/pgf/number format/.cd, fixed,/tikz/.cd},
        yticklabel style={font=\tiny,/pgf/number format/.cd, fixed,/tikz/.cd},
        x label style={at={(ticklabel cs:0.5, -5pt)},name={x label},anchor=north,font=\scriptsize},
        y label style={at={(ticklabel cs:0.5, -5pt)},name={y label},anchor=south,font=\scriptsize},
    },
    automatically generated symbolic/.style={
        height=100pt, width=185pt,
        xticklabel style={font=\small},
        yticklabel style={font=\small,/pgf/number format/.cd, fixed,/tikz/.cd},
        x label style={at={(ticklabel cs:0.5, -5pt)},name={x label},anchor=north,font=\footnotesize},
        y label style={at={(ticklabel cs:0.5, -5pt)},name={y label},anchor=south,font=\footnotesize},
    },
    first kind/.style={
		bar legend,
		legend columns=2,
    }
}
\tikzset{
    automatically generated plot/.style={
        /pgfplots/error bars/x dir=both, /pgfplots/error bars/y dir=both,
        /pgfplots/error bars/x explicit, /pgfplots/error bars/y explicit,
        /pgfplots/error bars/error bar style={ultra thin,solid},
        /tikz/mark options={solid},
    },
    automatically generated bar plot/.style={
        /pgfplots/error bars/y dir=both, /pgfplots/error bars/y explicit,
        /pgfplots/error bars/error bar style={line width=6pt,solid,black!40},
        /pgfplots/error bars/error mark=none,
    }
}

\def\xInterferencedxwithxRandomPermutationxtxt{}
\def\xInterferencedxwithxRandomPermutationx{0}
\def\xNoxinterferencextxt{"No interference"}
\def\xNoxinterferencex{1}

\begin{experimentfigure}
    \centering
    \begin{tikzpicture}[baseline,remember picture]
    \begin{axis}[
        automatically generated symbolic,
        width=15cm, height=5cm,
        xtick={1,2,3,4,5},
        xticklabels = {{All2All}, {RandomInvolution}, {Stencil von Neumann}, {Stencil Moore}, {All-reduce}},
        xticklabel style={align=center, font=\scriptsize, rotate=0},
        ybar,bar width=5.5pt,
        enlarge x limits=0.15,
        first kind,,
		legend style={at={(0.85,0.99)},anchor=north east,fill=white,fill opacity=0.8, text opacity=1},
        ymin=0,
		ymax=3.2e5,
        ymajorgrids=true, yminorgrids=true, xmajorgrids=true,
        mark options=solid, minor y tick num=4,
        ylabel={Cycles},
    ]
        \addplot[xxxxDiagonalsxxxxxbar] coordinates{
            (1,129103.60) -=(0, 0) +=(0, 79918.00)
            (2,80184.80) -=(0, 0) +=(0, 45832.00)
            (3,35669.60) -=(0, 0) +=(0, 11009.20)
            (4,76403.60) -=(0, 0) +=(0, 20674.40)
            (5,85675.60) -=(0, 0) +=(0, 47800.80)
        };\addlegendentry{\xxxxDiagonalsxxxxxtext}

        \addplot[xxxxRandomxSwitchxSelectionxxxxxbar] coordinates{
            (1,129184.80) -=(0, 0) +=(0, 112834.00)
            (2,80160.00) -=(0, 0) +=(0, 76098.40)
            (3,35502.00) -=(0, 0) +=(0, 14954.80)
            (4,78620.00) -=(0, 0) +=(0, 24509.60)
            (5,87768.80) -=(0, 0) +=(0, 47836.40)
        };\addlegendentry{\xxxxRandomxSwitchxSelectionxxxxxtext}

        \addplot[xxxxRVIIIxRIVxRIIxxxxxbar] coordinates{
            (1,129222.80) -=(0, 0) +=(0, 99275.60)
            (2,80072.00) -=(0, 0) +=(0, 83517.20)
            (3,35510.00) -=(0, 0) +=(0, 23986.80)
            (4,90471.60) -=(0, 0) +=(0, 27171.20)
            (5,87244.40) -=(0, 0) +=(0, 61645.20)
        };\addlegendentry{\xxxxRVIIIxRIVxRIIxxxxxtext}

        \addplot[xxxxLxtilesxxxxxbar] coordinates{
            (1,129185.60) -=(0, 0) +=(0, 125368.00)
            (2,90712.00) -=(0, 0) +=(0, 92580.80)
            (3,37518.00) -=(0, 0) +=(0, 28415.20)
            (4,94766.00) -=(0, 0) +=(0, 26459.60)
            (5,104558.80) -=(0, 0) +=(0, 49740.80)
        };\addlegendentry{\xxxxLxtilesxxxxxtext}

        \addplot[xxxxRowsxxxxxbar] coordinates{
            (1,133096.40) -=(0, 0) +=(0, 132629.60)
            (2,169692.00) -=(0, 0) +=(0, 35402.80)
            (3,47380.40) -=(0, 0) +=(0, 29513.60)
            (4,94840.40) -=(0, 0) +=(0, 41707.60)
            (5,143244.80) -=(0, 0) +=(0, 31640.80)
        };\addlegendentry{\xxxxRowsxxxxxtext}

        \addplot[xxxxRandomxxxxxbar] coordinates{
            (1,129112.00) -=(0, 0) +=(0, 86404.00)
            (2,79955.60) -=(0, 0) +=(0, 215310.80)
            (3,34901.60) -=(0, 0) +=(0, 38676.80)
            (4,75774.80) -=(0, 0) +=(0, 74203.60)
            (5,84500.80) -=(0, 0) +=(0, 124321.60)
        };\addlegendentry{\xxxxRandomxxxxxtext}

        \addplot[xxxxRRxswitchesxxxxxbar] coordinates{
            (1,129068.80) -=(0, 0) +=(0, 88731.20)
            (2,79912.00) -=(0, 0) +=(0, 223706.00)
            (3,32879.20) -=(0, 0) +=(0, 39923.20)
            (4,69309.60) -=(0, 0) +=(0, 73399.60)
            (5,74734.40) -=(0, 0) +=(0, 143223.20)
        };\addlegendentry{\xxxxRRxswitchesxxxxxtext}

    \end{axis}
    \end{tikzpicture}
    \caption{Makespan for different 64-process kernels. Background traffic is a random permutation. The light-colored bars show the time running isolated and the dark part shows the extra time taken due to the interference.}
    \label{fig:random-kernel-inter-no-VC}
\end{experimentfigure}

\egroup

The application escalation study has been done considering 64, 128 and 256-process applications. For brevity only the case of 64 processes is shown in Figure~\ref{fig:omni_64_random}, since it presents the most interesting situations due to the differences observed. However, all the experiments are also summarized in Table~\ref{tab:tablespeedups2}.
The best strategies are close to Diagonal allocation across all configurations. For this reason, values in the table are normalized with respect to it. Full Spread and Random Endpoint occasionally achieve slightly higher mean speedups (up to about 1.05) at 50\% load, but the differences are small and tend to disappear at 100\% load. 	Random Switch is close to Diagonal but Rectangular is clearly worse. Row and L-shape consistently exhibit mean slowdowns, particularly for 64–128 ranks and at 50\% occupancy.

In summary, in these experiments, Full Spread, Random Endpoint, Diagonal and Random Switch selection functions form a group of high-quality strategies with similar behaviour across applications and loads, whereas Row, L-shape Tessellation and Rectangular selection functions are clearly less favorable in terms of average execution time.
Lastly, with bigger applications, the differences are less pronounced.

	\begin{table}[htbp]
		\centering
		\scriptsize
		\caption{Average performance of each strategy normalized to the performance of Diagonal Selection. Employing \textbf{Omni-WAR routing}. Including values for a half-occupied and a completely-filled system.}
		\begin{tabular}{lcccccc}
			\toprule
			& \multicolumn{2}{c}{64 ranks}
			& \multicolumn{2}{c}{128 ranks}
			& \multicolumn{2}{c}{256 ranks} \\
			\cmidrule(lr){2-3}\cmidrule(lr){4-5}\cmidrule(lr){6-7}
			& 50\% & 100\%
			& 50\% & 100\%
			& 50\% & 100\% \\
			\midrule
			Diagonal   & 1.00 & 1.00 & 1.00 & 1.00 & 1.00 & 1.00 \\
			R. Switch  & 0.99 & 0.93 & 0.95 & 0.96 & 0.98 & 0.97 \\
			Rect.   & 0.88 & 0.88 & 0.89 & 0.99 & - & - \\
			L-shape  & 0.77 & 0.88 & 0.87 & 0.97 & 0.91 & 0.97 \\
			Row & 0.74 & 0.93 & 0.77 & 0.94 & 0.87 & 0.96 \\
			R. Endpoint   & 1.02 & 1.00 & 0.99 & 1.03 & 1.01 & 1.02 \\
			Full S.   & 1.05 & 1.00 & 1.02 & 1.03 & 1.03 & 1.02 \\
			\bottomrule
		\end{tabular}\label{tab:tablespeedups2}
	\end{table}

	\begin{table}[htbp]
		\centering
		\small
		\caption{Average slowdown in the interference analysis scenario of each allocation function normalized to the performance of Diagonal Selection.}
		\begin{tabular}{lcc}
			\toprule
			& 64 ranks & 128 ranks \\
			\midrule
			Diagonal Selection   & 1.00 & 1.00 \\
			Random Switch Selection   & 0.90 & 0.98 \\
			Rectangular Tessellation  & 0.84 & 0.93 \\
			L-shape Tessellation      & 0.78 & 0.96 \\
			Row Selection             & 0.70 & 0.93 \\
			Random Endpoint Selection   & 0.66 & 0.80 \\
			Full Spread               & 0.66 & 0.76 \\
			\bottomrule
		\end{tabular}\label{tab:tablespeedups1}
	\end{table}

\subsubsection{Interference analysis}\label{subsec:interference}

	Figure~\ref{fig:random-kernel-inter-no-VC} reports the per-application interference experiments, and Table~\ref{tab:tablespeedups1} condenses them into the mean slowdown of each strategy relative to Diagonal for 64 and 128 ranks.
	The observed differences indicate that the choice of mapping becomes notable, and that Diagonal is the most robust strategy on average, with Random Switch as a reasonable alternative.
	The robustness of a partition must be understood as the ability to remain as immune as possible to external traffic interferences.

	For 64 ranks, Diagonal is substantially faster than Rectangular, Random Endpoint, Full Spread, L-shape and Row allocation functions, while Random Switch Selection is closer.
At 128 ranks, differences are smaller but the same trend holds, with Random Endpoint and Full Spread occupying the last positions.
In this respect, in Section~\ref{sec:vcs} it is shown how the addition of costly VCs can mitigate the effects of interference in both Full Spread and Random Endpoint, improving their performance and outperforming the other strategies.
Simulations with 256 ranks were also conducted but are not reported here for brevity, as they yielded similar conclusions to the other cases.

	\subsection{Fabric partitioning}\label{sec:vcs}

	\bgroup
\pgfplotsset{compat=newest}
\makeatletter
\tikzset{nomorepostaction/.code=\let\tikz@postactions\pgfutil@empty}
\makeatother
\pgfplotsset{minor grid style={dashed,very thin, color=blue!15}}
\pgfplotsset{major grid style={very thin, color=black!30}}
\pgfplotsset{
    automatically generated axis/.style={
       height=105pt, width=174pt, scaled ticks=false,
       xticklabel style={font=\tiny,/pgf/number format/.cd, fixed,/tikz/.cd},
       yticklabel style={font=\tiny,/pgf/number format/.cd, fixed,/tikz/.cd},
       x label style={at={(ticklabel cs:0.5, -5pt)},name={x label},anchor=north,font=\scriptsize},
       y label style={at={(ticklabel cs:0.5, -5pt)},name={y label},anchor=south,font=\scriptsize},
    },
    automatically generated symbolic/.style={
       height=125pt, width=245pt,%
       xticklabel style={font=\small},
       yticklabel style={font=\small,/pgf/number format/.cd, fixed,/tikz/.cd},
       x label style={at={(ticklabel cs:0.5, -5pt)},name={x label},anchor=north,font=\footnotesize},
       y label style={at={(ticklabel cs:0.5, -5pt)},name={y label},anchor=south,font=\footnotesize},
    },
    first kind/.style={
	   bar legend,
	   legend columns=3,
    }
}
\tikzset{
    automatically generated plot/.style={
       /pgfplots/error bars/x dir=both, /pgfplots/error bars/y dir=both,
       /pgfplots/error bars/x explicit, /pgfplots/error bars/y explicit,
       /pgfplots/error bars/error bar style={ultra thin,solid},
       /tikz/mark options={solid},
    },
    automatically generated bar plot/.style={
       /pgfplots/error bars/y dir=both, /pgfplots/error bars/y explicit,
       /pgfplots/error bars/error bar style={line width=7pt,solid,black!40},
       /pgfplots/error bars/error mark=none,
    }
}

\def\xInterxwithxrandomxswitchxpermxtxt{"Inter with random switch perm"}
\def\xInterxwithxuniformxtxt{"Inter with uniform"}
\def\xInterferencedxwithxRandomPermutationxtxt{"Interferenced with RandomPermutation"}
\def\xInterferencedxwithxRandomPermutationx{0}
\def\xNoxInterferredxtxt{"No Interferred"}
\def\xNoxInterferredx{3}

\def\xxxxDiagonalsxxxxxtext{Diagonal Sel.}
\def\xxxxRandomxSwitchxSelectionxxxxxtext{Random Switch Sel.}
\def\xxxxRVIIIxRIVxRIIxxxxxtext{Rectangular Tes.}
\def\xxxxLxtilesxxxxxtext{L-shape Tes.}
\def\xxxxRandomxxxxxtext{Random Endpoint Sel.}

\begin{experimentfiguressmall}
    \centering
    \pgfplotslegendfromname{leyenda-random-inter-con-VC}\\
    \begin{tikzpicture}[baseline,remember picture]
    \begin{axis}[
       automatically generated symbolic,
       xtick={0, 1}, xticklabels = {{Uniform}, {Random Switch Perm}},
       ybar,bar width=8pt,enlarge x limits=0.6,
       first kind,,
        legend to name=leyenda-random-inter-con-VC,
       ymin=0,
       ymajorgrids=true, yminorgrids=true, xmajorgrids=true,
       mark options=solid, minor y tick num=4,
       ylabel={Cycles},
    ]
        \addplot[xxxxDiagonalsxxxxxbar] coordinates{  (\xInterferencedxwithxRandomPermutationx,92988.80) -=(0, 0) +=(0, 19295.20) (1,80116.40) -=(0, 0) +=(0, 104155.20)};\addlegendentry{\xxxxDiagonalsxxxxxtext}
        \addplot[xxxxRandomxSwitchxSelectionxxxxxbar] coordinates{  (\xInterferencedxwithxRandomPermutationx,94876.80) -=(0, 0) +=(0, 25433.20) (1,92193.20) -=(0, 0) +=(0, 168729.20)};\addlegendentry{\xxxxRandomxSwitchxSelectionxxxxxtext}
        \addplot[xxxxRVIIIxRIVxRIIxxxxxbar] coordinates{  (\xInterferencedxwithxRandomPermutationx,105497.60) -=(0, 0) +=(0, 43751.20) (1,110809.60) -=(0, 0) +=(0, 221024.80)};\addlegendentry{\xxxxRVIIIxRIVxRIIxxxxxtext}
        \addplot[xxxxLxtilesxxxxxbar] coordinates{  (\xInterferencedxwithxRandomPermutationx,111478.00) -=(0, 0) +=(0, 30076.00) (1,117831.20) -=(0, 0) +=(0, 192150.40)};\addlegendentry{\xxxxLxtilesxxxxxtext}
        \addplot[xxxxRowsxxxxxbar] coordinates{  (\xInterferencedxwithxRandomPermutationx,112452.00) -=(0, 0) +=(0, 60758.00) (1,148258.40) -=(0, 0) +=(0, 143048.40)};\addlegendentry{\xxxxRowsxxxxxtext}
        \addplot[xxxxRandomxxxxxbar] coordinates{  (\xInterferencedxwithxRandomPermutationx,92352.00) -=(0, 0) +=(0, 369.20) (1,79962.40) -=(0, 0) +=(0, 20548.40)};\addlegendentry{\xxxxRandomxxxxxtext}
        \addplot[xxxxRRxswitchesxxxxxbar] coordinates{  (\xInterferencedxwithxRandomPermutationx,93045.60) -=(0, 0) +=(0, 984.00) (1,79880.00) -=(0, 0) +=(0, 1193.20)};\addlegendentry{\xxxxRRxswitchesxxxxxtext}
    \end{axis}
    \end{tikzpicture}
    \caption{Makespan of 64-process kernels providing a different VC set to the background traffic. The dark part shows the extra time taken due to the interference over running isolated.}
    \label{fig:random-inter-con-VC}
\end{experimentfiguressmall}

\egroup

	\bgroup
\pgfplotsset{compat=newest}
\makeatletter
\tikzset{nomorepostaction/.code=\let\tikz@postactions\pgfutil@empty}
\makeatother
\pgfplotsset{minor grid style={dashed,very thin, color=blue!15}}
\pgfplotsset{major grid style={very thin, color=black!30}}
\pgfplotsset{
    automatically generated axis/.style={
        height=105pt, width=174pt, scaled ticks=false,
        xticklabel style={font=\tiny,/pgf/number format/.cd, fixed,/tikz/.cd},
        yticklabel style={font=\tiny,/pgf/number format/.cd, fixed,/tikz/.cd},
        x label style={at={(ticklabel cs:0.5, -5pt)},name={x label},anchor=north,font=\scriptsize},
        y label style={at={(ticklabel cs:0.5, -5pt)},name={y label},anchor=south,font=\scriptsize},
    },
    automatically generated symbolic/.style={
        height=125pt, width=185pt,
        xticklabel style={font=\small},
        yticklabel style={font=\small,/pgf/number format/.cd, fixed,/tikz/.cd},
        x label style={at={(ticklabel cs:0.5, -5pt)},name={x label},anchor=north,font=\footnotesize},
        y label style={at={(ticklabel cs:0.5, -5pt)},name={y label},anchor=south,font=\footnotesize},
    },
    first kind/.style={
		bar legend,legend columns=2,
    }
}
\tikzset{
    automatically generated plot/.style={
        /pgfplots/error bars/x dir=both, /pgfplots/error bars/y dir=both,
        /pgfplots/error bars/x explicit, /pgfplots/error bars/y explicit,
        /pgfplots/error bars/error bar style={ultra thin,solid},
        /tikz/mark options={solid},
    },
    automatically generated bar plot/.style={
        /pgfplots/error bars/y dir=both, /pgfplots/error bars/y explicit,
        /pgfplots/error bars/error bar style={line width=6pt,solid,black!40},
        /pgfplots/error bars/error mark=none,
    }
}

\def\xInterferencedxwithxRandomPermutationxtxt{}
\def\xInterferencedxwithxRandomPermutationx{0}
\def\xNoxinterferencextxt{"No interference"}
\def\xNoxinterferencex{1}

\begin{experimentfigure}
    \centering
    \begin{tikzpicture}[baseline,remember picture]
    \begin{axis}[
        automatically generated symbolic,
        width=15cm, height=5cm,
        xtick={1,2,3,4,5},
        xticklabels = {{All2All}, {RandomInvolution}, {Stencil von Neumann}, {Stencil Moore}, {All-reduce}},
        xticklabel style={align=center, font=\scriptsize, rotate=0},
        ybar,bar width=5pt,
        enlarge x limits=0.15,
        first kind,,
		legend style={at={(0.82,0.99)},anchor=north east,fill=white,fill opacity=0.8, text opacity=1},
        ymin=0,
		ymax=2.6e5,
        ymajorgrids=true, yminorgrids=true, xmajorgrids=true,
        mark options=solid, minor y tick num=4,
        ylabel={Cycles},
    ]
        \addplot[xxxxDiagonalsxxxxxbar] coordinates{
            (1,129103.60) -=(0, 0) +=(0, 59758.80)
            (2,80184.80) -=(0, 0) +=(0, 56506.80)
            (3,35669.60) -=(0, 0) +=(0, 8543.20)
            (4,76403.60) -=(0, 0) +=(0, 10876.00)
            (5,85675.60) -=(0, 0) +=(0, 49659.60)
        };\addlegendentry{\xxxxDiagonalsxxxxxtext}

        \addplot[xxxxRandomxSwitchxSelectionxxxxxbar] coordinates{
            (1,129184.80) -=(0, 0) +=(0, 90610.00)
            (2,80160.00) -=(0, 0) +=(0, 68622.40)
            (3,35502.00) -=(0, 0) +=(0, 12280.80)
            (4,78620.00) -=(0, 0) +=(0, 8760.40)
            (5,87768.80) -=(0, 0) +=(0, 48640.40)
        };\addlegendentry{\xxxxRandomxSwitchxSelectionxxxxxtext}

        \addplot[xxxxRVIIIxRIVxRIIxxxxxbar] coordinates{
            (1,129222.80) -=(0, 0) +=(0, 110962.00)
            (2,80072.00) -=(0, 0) +=(0, 87456.80)
            (3,35510.00) -=(0, 0) +=(0, 22290.80)
            (4,90471.60) -=(0, 0) +=(0, 22878.80)
            (5,87244.40) -=(0, 0) +=(0, 61169.60)
        };\addlegendentry{\xxxxRVIIIxRIVxRIIxxxxxtext}

        \addplot[xxxxLxtilesxxxxxbar] coordinates{
            (1,129185.60) -=(0, 0) +=(0, 119372.40)
            (2,90712.00) -=(0, 0) +=(0, 101218.40)
            (3,37518.00) -=(0, 0) +=(0, 27500.80)
            (4,94766.00) -=(0, 0) +=(0, 23465.60)
            (5,104558.80) -=(0, 0) +=(0, 56331.60)
        };\addlegendentry{\xxxxLxtilesxxxxxtext}

        \addplot[xxxxRowsxxxxxbar] coordinates{
            (1,133096.40) -=(0, 0) +=(0, 116932.40)
            (2,169692.00) -=(0, 0) +=(0, 48474.00)
            (3,47380.40) -=(0, 0) +=(0, 22174.40)
            (4,94840.40) -=(0, 0) +=(0, 37230.40)
            (5,143244.80) -=(0, 0) +=(0, 43456.80)
        };\addlegendentry{\xxxxRowsxxxxxtext}

        \addplot[xxxxRandomxxxxxbar] coordinates{
            (1,129112.00) -=(0, 0) +=(0, 5249.60)
            (2,79955.60) -=(0, 0) +=(0, 13777.60)
            (3,34901.60) -=(0, 0) +=(0, 0.00)
            (4,75774.80) -=(0, 0) +=(0, 0.00)
            (5,84500.80) -=(0, 0) +=(0, 22162.80)
        };\addlegendentry{\xxxxRandomxxxxxtext}

        \addplot[xxxxRRxswitchesxxxxxbar] coordinates{
            (1,129068.80) -=(0, 0) +=(0, 295.60)
            (2,79912.00) -=(0, 0) +=(0, 751.60)
            (3,32879.20) -=(0, 0) +=(0, 402.40)
            (4,69309.60) -=(0, 0) +=(0, 2455.60)
            (5,74734.40) -=(0, 0) +=(0, 13846.00)
        };\addlegendentry{\xxxxRRxswitchesxxxxxtext}

    \end{axis}
    \end{tikzpicture}
    \caption{Makespan of 64-process kernels providing a different VC set to the background traffic. The light-colored bars show the time running isolated and the dark part shows the extra time taken due to the interference.}
    \label{fig:random-kernel-interference-with-VC}
\end{experimentfigure}

\egroup

	Figures~\ref{fig:random-inter-con-VC} and \ref{fig:random-kernel-interference-with-VC} show interference analysis results with more virtual channels, for both basic traffic and kernel traffic.
	The addition of virtual channels has a significant impact on the performance of the different strategies, particularly for those that are more sensitive to HoL blocking, such as Full Spread and Random Endpoint Selection.
	With more virtual channels, these strategies are able to separate different traffic flows and reduce contention for resources, leading to improved performance.
	Full Spread and Random Endpoint outperform Diagonal, which was previously the best strategy.
	This highlights the importance of considering virtual channel configurations when evaluating resource allocation strategies in HyperX networks.
Nevertheless, the number of virtual channels must be kept as small as possible due to cost and power constraints.

	\section{Lessons Learned}\label{sec:outcomes}

	In this section lessons learned from the experimentation are established and explained in the light of the topological characteristics of the partitions.

	\begin{lesson}
		Lesson 1: The HyperX topology demands for new resource allocation strategies.
	\end{lesson}

	The experimental results confirm that the most relevant performance metric for a resource allocation strategy is strongly topology-dependent, and that heuristics effective in Tori, Fat-trees, and Dragonflies do not transfer directly to HyperX.

	As discussed in Section~\ref{sec:background}, allocation strategies in Tori are designed to minimize dilation.
	If dilation were also the dominant factor in HyperX, Row allocation—in which all endpoints are at distance one—would be expected to perform best.
	However, experiments show several cases in which Row exhibits the worst performance, most notably for Random Involution and All-reduce patterns.
	In the same way, L-shapes, which also provide low average distance, yield poor results in most scenarios.
	These observations indicate that, unlike in Tori, minimizing dilation alone is not sufficient in HyperX.

	A similar deviation is observed when comparing HyperX versus Fat-tree networks. In Fat-trees, convexity of allocated partitions is a key property for job placement, as it helps isolate partition traffic by exploiting the hierarchical network structure.
	In HyperX, by contrast, non-convex partitions can provide better performance: for example, Diagonal, which is non-convex, consistently outperforms convex strategies such as Row and Rectangular in most scenarios, as shown experimentally. Furthermore, when interference is evaluated, the Row strategy—despite allocating convex partitions—fails to isolate traffic effectively due to non-minimal routes going through target application partition, leading to significant performance degradation.

	Lastly, although resource allocation for Dragonflies aims to maximize bandwidth, its strategies do not carry over directly.
	As recalled in Section~\ref{sec:background}, Dragonfly allocation policies typically either pack an application within a single group or spread it across the network using as few endpoints per group as possible.
	The natural analogues in HyperX would be to place small applications within a single row (Row allocation) and to use a strategy such as Full Spread for large applications.
	However, for applications of size 64—i.e., those that would fit in a single “group” (row or column)—Row performs clearly poorly.
	In addition, Full spread alone leads to a bad performance in interference scenarios limited to 4VCs, as shown in Figure~\ref{fig:omni-basic-traffic-interference-no-vc}.

	Altogether, these  results indicate that, while dilation is the central metric in Tori, convexity in Fat-trees, and bandwidth in Dragonflies, HyperX requires dedicated allocation strategies and cannot rely on a direct reuse of those devised for other topologies.

	\begin{lesson}
		Lesson 2: Partition Bandwidth emerges as a key performance indicator.
	\end{lesson}

	Bandwidth availability and its distribution across the topology appear as decisive aspects.
	In low interference scenarios, Full Spread, Random Endpoint, Diagonal and Random Switch strategies are, in this order, the schemes that provide the highest empirical throughput to the allocated application, as Table~\ref{tab:tablespeedups2} shows.
	The measured values are quite close but they are clearly better than the others.
	This perfectly correlates with the values provided by the theoretical partition bandwidth metric, $\PB$, introduced in Subsection \ref{subsec:bandwidth_analysis}.
	In the same order and particularized for $n=8$, Full Spread, Random Endpoint, Diagonal, and Random Switch have a $\PB$ of 8.0, 5.1, 2.0, and 1.3, respectively.
	The other three strategies, Rectangular, L-shape and Row, have $\PB \leq 1$.

	The impact of the partition bandwidth is more pronounced at lower occupancy levels, where contention is reduced and the benefits of higher throughput allocations can be fully exploited.
	For instance, the maximum speedup of Diagonal over other strategies across all applications is 2.16 with one application allocated and 2.30 with half the endpoints allocated, under RandomInvolution traffic, and 1.30 with all endpoints allocated under All2All traffic, as shown in Figure~\ref{fig:omni_64_random}.

	All other experiments show similar trends, confirming that partition bandwidth availability is a key factor in determining the effectiveness of a resource allocation strategy in HyperX networks.

	\begin{lesson}
		Lesson 3: Under certain demanding load conditions, strategies that do not preserve switch locality result in lower performance due to interference and HoL blocking.
	\end{lesson}

	Partition Bandwidth alone is not capable of estimating the performance of each allocation strategy, especially in scenarios under high interference.
	It is the combination of partition bandwidth along with the HoL blocking and interference patterns that determines the performance of each strategy.

	While all allocation strategies benefit from low system occupancy, the impact of interferences becomes evident as the load increases and competition for switches and links is intense.
	Full Spread and Random Endpoint create partitions that do not preserve switch-locality, this is, they share switches with other concurrent workloads.
	Consequently, applications allocated via these methods provoke higher interferences than using partitions that preserve switch-locality, resulting in lower performance.
	In contrast, all other strategies do preserve switch-locality, being Diagonal and Random Switch the ones exhibiting better performance, as shown in Figure~\ref{fig:random-kernel-inter-no-VC}.

	\begin{lesson}
		Lesson 4: If fabric partitioning is used to isolate application traffic, strategies which maximize partition bandwidth are the best.
	\end{lesson}

 Partitions such as Full Spread and Random Endpoint achieve the highest PB; however, they often suffer from significant HoL blocking because they do not preserve switch locality, spreading applications across the entire network. By implementing virtual channels to isolate flows, this HoL blocking is mitigated. This allows the high bandwidth potential of these strategies to be fully exploited, resulting in superior performance. In contrast, strategies like Row do not see similar performance gains from virtual channels, as their low partition bandwidth remains its primary bottleneck.

	\section{Conclusions}\label{sec:discussion}

	HyperX networks are presented as efficient alternatives to other topologies based on complete-graph constructions, such as Dragonfly or Dragonfly+. These networks stand out for their favorable distance properties, good scalability, and notable cost–performance ratio. Regarding their deployment as system networks in the market, several works have explored how to address practical challenges in these topologies, including routing \cite{Kim_omni} and fault tolerance \cite{SurePath}. In the present study, the authors focus on the problem of partitioning and resource allocation, given its potential impact on performance, as previously observed in Torus and Dragonfly-type networks.

	To this end, different resource allocation functions for HyperX networks have been modeled, taking into account algebraic, geometric, and stochastic aspects. For these functions, several relevant properties have been characterized, namely dilation, convexity, and the resulting local bandwidth. In addition, an extensive experimental evaluation of the different strategies has been conducted under two increasingly realistic scenarios: from the scaling behavior of a single application to a setting in which interference among multiple applications running simultaneously on the system is evaluated.

	Considering these theoretical and experimental results, the article provides a set of lessons learned that serve as guidelines for the implementation of resource allocation policies in HyperX based systems. Although the HyperX topology proves to be largely immune to suboptimal decisions in this respect, choosing an effective routing algorithm or selecting the most stable allocation policy can still yield performance gains of up to $2.30 \times$.

	The article focuses exclusively on partitioning 2D HyperX networks. However, 3D HyperX networks can also benefit from this study, as our methodology should extend naturally to higher dimensions. For example, it is clear that functions based on algebraic criteria—such as linear functions—or those based on random permutations can be defined without difficulty in $q$ dimensions.

    In summary, the study presented in this article provides yet another reason for HyperX networks to be regarded as excellent alternatives to be used in forthcoming HPC systems.

\section*{Acknowledgment}

This work has been supported by the Spanish Ministry of Science and Innovation under contracts  PID2022-136454NB-C21 and RYC2021-033959-I, the European HiPEAC Network of Excellence, and the Barcelona Supercomputing Center under contract CONSER02023011NG.

	\bibliographystyle{plain}
	\bibliography{main}

	\clearpage

\end{document}